\documentclass[]{interact}

\usepackage{epstopdf}
\usepackage{hyperref}
\usepackage[numbers,sort&compress]{natbib}
\bibpunct[, ]{[}{]}{,}{n}{,}{,}
\renewcommand\bibfont{\fontsize{10}{12}\selectfont}
\makeatletter
\def\NAT@def@citea{\def\@citea{\NAT@separator}}
\makeatother

\theoremstyle{plain}

\theoremstyle{definition}

\theoremstyle{remark}

\usepackage{siunitx}
\usepackage{placeins}
\usepackage{subcaption}

\usepackage{physics}
\usepackage{amsmath}
\usepackage{tikz}
\usepackage{mathdots}
\usepackage{yhmath}
\usepackage{cancel}
\usepackage{color}
\usepackage{siunitx}
\usepackage{array}
\usepackage{multirow}
\usepackage{amssymb}
\usepackage{gensymb}
\usepackage{tabularx}
\usepackage{extarrows}
\usepackage{booktabs}
\usetikzlibrary{fadings}
\usetikzlibrary{patterns}
\usetikzlibrary{shadows.blur}
\usetikzlibrary{shapes}
\usetikzlibrary{arrows,backgrounds}

\begin{document}

\title{
The role of fluid friction in streamer formation and biofilm growth\\
}
\author{
\name{Cornelius Wittig\textsuperscript{1}, Michael Wagner\textsuperscript{2}, Romain Vallon \textsuperscript{1}, Thomas Crouzier\textsuperscript{3}, Wouter van der Wijngaart\textsuperscript{4}, Harald Horn\textsuperscript{5}, and Shervin Bagheri\textsuperscript{1}}
\affil{\textsuperscript{1}FLOW Center, Dept. of Engineering Mechanics, KTH, Stockholm SE-100 44, Sweden\\ 
\textsuperscript{2}Karlsruhe Institute of Technology, Institute of Biological Interfaces (IBG-1), Eggenstein-Leopoldshafen, Germany\\ 
\textsuperscript{3}DTU, Dept. of Health Technology, DK-2800 Kongens Lyngby, Denmark\\
\textsuperscript{4}Division of Micro and Nanosystems, Dept. of Intelligent Systems, KTH, Stockholm SE-100 44, Sweden\\ \textsuperscript{5}Karlsruhe Institute of Technology, Water Chemistry and Water Technology, Engler-Bunte-Institut, Karlsruhe, Germany}
}

\maketitle

\begin{abstract}
\textit{Bacillus subtilis} biofilms were grown in laminar channel flow at wall shear stress spanning one order of magnitude ($\tau_w = \SI{0.068}{\pascal}$ to $\tau_w = \SI{0.67}{\pascal}$). We monitor, non-invasively, the evolution of the three-dimensional distribution of biofilm over seven days using optical coherence tomography (OCT). The obtained biofilms consist of many microcolonies where the characteristic colony has a base structure in the form of a leaning pillar and a streamer in the form of a thin filament that originates near the tip of the pillar. While the shape, size and distribution of these microcolonies depend on the imposed shear stress, the same structural features appear consistently for all shear stress values. The formation of streamers seems to occur after the development of a base structure, suggesting that the latter induces a curved secondary flow that triggers the formation of the streamers. Moreover, we observe that the biofilm volume grows approximately linearly over seven days for all the shear stress values, with a growth rate that is inversely proportional to the wall shear stress. 
We develop a simple model of friction-limited growth, which agrees with the experimental observations. The model provides physical insight into growth mechanisms and can be used to develop accurate continuum models of bacterial biofilm growth.

\end{abstract}

\begin{keywords}
OCT; biofilm; wall shear stress; growth rate; streamers; mesoscale
\end{keywords}

\section{Introduction}
Bacteria suspended in a matrix of extracellular polymeric substances (EPS), also known as biofilms, grow on nearly all engineered surfaces in aquatic environments. In a biofilm, EPS serves to protect the bacteria from biological, chemical, and mechanical stress \citep{Flemming.2010, Shaw.2004}. The resulting biofilm may be described as a growing, viscoelastic material \citep{Jana.2020}.
In natural settings and most applications, the external environment for biofilms is a flowing fluid, which serves as a means of delivery of nutrients, oxygen and other substances necessary for the biofilm to survive. However, the flow also imposes forces on the biofilm.
These forces shape the biofilm through processes such as erosion, sloughing, and other fluid-structure interactions.
The precise role that shear forces have on biofilm growth is not fully understood. 
One particularly challenging aspect in understanding how a shear flow influences biofilms -- regardless of flow regime and type of biofilm  -- is the decoupling of the effects caused by fluid shear stress, nutrient transport by the flow, and the biological response to mechanical cues. 
However, for most species, it has been observed that higher shear stresses tend to inhibit the growth of biofilms \citep{Chun.2022, Thomen.2017} and typically lead to the formation of more compact biofilms \citep{Simoes.2007, Wagner.2010}. 
Shear stress also influences the morphology of the biofilm, with perhaps the most distinct feature being streamers, which are relatively thin filamentous structures. 
Streamers have been extensively reported in turbulent flows \citep{Stoodley.1998,Hartenberger.2020} and in microfluidic channels \citep{Rusconi.2010,Drescher.2013,Kurz.2022}.
The underlying reasons of how and why these streamers appear, and their interaction with flowing fluid remain unclear. One hypothesis \citep{Rusconi.2010,Carpio.2016} is related to secondary motion promoting steamer development, i.e. localised re-circulation zones resulting in curved streamlines and accumulation of biomass. Indeed, turbulence, corners and other geometric defects cause secondary motions with varying intensity. In contrast, except for the side wall, a laminar flow in a channel flow does not contain any secondary flows. However, Chun et al. \cite{Chun.2022} found streamers in a laminar channel flow at a shear stress above $\tau_w = \SI{5}{\pascal}$ for \textit{Cobetia marina} and \textit{Pseudonomas aeruginosa} biofilms. In this work, we will explore whether the observed streamers are related to secondary flow induced by the biofilm itself or to other potential mechanisms.
We observe the consistent existence of streamers on biofilm structures that grow on smooth flat walls over an order of magnitude of shear stress and several days. In fact, our biofilms consist of a forest of microcolonies with streamers that align with the bulk flow. 

Another aspect is that there is no general model for friction-limited biofilm growth. 
This is in contrast to nutrient-limited models, where advective and diffusive transport mechanisms coupled with substrate uptake kinetics (e.g. via the Monod equation) have been shown to quantitatively explain experimental observations \citep{Horn.1997, Picioreanu.1998, MartinezCalvo.2022}. To resist detachment, biofilm absorbs external stress partly by behaving as a viscous fluid and partly via elastic deformation \citep{Peterson.2015}. Biofilms are capable of adapting to varying external forcing by adjusting their viscoelastic properties, for example by an increase in the extracellular polymer production by the cells.

Previous studies studies have either investigated biofilm growth over time at a single value of shear stress \citep{Horn.1997, Bakke.2001, Gierl.2020} or as end-point measurements for varying wall-shear stress values \citep{Paul.2012, Chun.2022}. A time-resolved measurement of the dependency of biofilm growth on shear stress is, however, missing.
The accumulation of biofilm at one level of wall shear stress has been investigated in several studies.
Horn and Hempel \citep{Horn.1997} observed an almost linear accumulation of biofilm in a circular pipe over 108 days, whereas Bakke et al. \citep{Bakke.2001} observed a similar trend in square ducts over 13 days. Gierl et al. \citep{Gierl.2020} introduced a robotic platform to monitor biofilm growth in multiple flow cells and confirmed these observations for \textit{Bacillus subtilis} over six days.
A multitude of end-point studies has shown that an increase in the wall shear stress impedes the growth of biofilm. Paul et al. \citep{Paul.2012} reported an exponential decline in the thickness of mature biofilms with increasing wall shear stress ($\tau_w=\SI{0.3}{\pascal}$ to $\tau_w=\SI{13}{\pascal}$). Chun et al. \citep{Chun.2022} investigated the very early stages of biofilm growth (4 hours) and showed for \textit{Cobetia marina} and \textit{Pseudonomas aeruginosa} that biofilm thickness, mass and coverage decreases linearly with shear stress ($\tau_w=\SI{0.2}{\pascal}$ to $\tau_w=\SI{5.6}{\pascal}$).
Here, building on the robotic platform presented by Gierl et al. \citep{Gierl.2020}, we present an experimental setup allowing the longitudinal study of biofilm growth under different shear stress conditions. We use this system to investigate the growth of biofilms formed by the same strain of \textit{Bacillus subtilis} under different levels of wall shear stress over seven days.
Three-dimensional scans of these biofilms are acquired in situ using optical coherence tomography (OCT).

OCT is an interferometric measurement method that has proven useful for examining biofilm and does not require staining of the sample. OCT is capable of penetrating biofilm much deeper than typical fluorescent microscopy methods, allowing the detection of voids within or beneath the biofilm. Measurements can be taken in situ, without altering the sample. This enables continued observation of the development of a biofilm over an extended period of time. With a voxel size of several microns along each dimension, and a maximum field of view on the order of $\SI{1}{\centi\metre\squared}$, OCT enables measurements of mesoscopic biofilm structures that are too large for confocal laser scanning microscopy.
Many different thresholding methods, often Otsu's method \citep{Otsu.1979}, are used to binarise the OCT scans \citep{Wagner.2017}. These methods struggle to separate the biofilm from the bulk if no bimodal distribution exists in the intensity histogram.
Here, we introduce a new thresholding algorithm based on the characteristics of the histogram of a typical OCT scan containing a small amount of biofilm.

By monitoring the mesoscopic development of biofilms over an extended period of time, we show how the wall shear stress influences the biofilm morphology. We find that the biofilm is a collection of many microcolonies that are shaped as tilted pillars with an attached streamer. Moreover, we develop a model of friction-limited growth that agrees with our experimental observation and that provides an explicit dependence of the biovolume on the wall shear stress and time. This model may be valuable in more advanced models for predicting the accumulation of biofilm on a surface, which is essential for many applications. In particular, 
the prediction of the drag increase of fouled ship hulls has been a subject of intensive research for many years. Previous studies have investigated the drag production by (patchy) biofilms \citep{Hartenberger.2020, Snowdon.2022, Murphy.2022}, and biofilm streamers \citep{Stoodley.1998, Taherzadeh.2010}. The fuel costs of fouling are considerable. For example, slime fouling can increase the required shaft power of a ship by up to $\SI{18}{\percent}$ \citep{Schultz.2011}.
It is our hope that this work will contribute to better models for predicting the early stages of biofouling.

\section{Materials and methods}

\begin{table}
    \centering
    \begin{tabular}{llcccccc}
    \toprule
    Case     &        & 1    & 2 & 3 & 4 & 5 & 6 \\
    \midrule
    $H$ & $[\si{mm}]$ & 2 & $\sqrt{2}$ & 2 & 1 & $\sqrt{2}$ & 1 \\
    $D_H$ & $[\si{mm}]$ & 3.63 & 2.64 & 3.63 & 2.64 & 1.9 & 1.9\\
    $Q$ & $[\si{\milli\litre\per\second}]$  & 1 & 1 & 2.63 & 1 & 2.63 & 2.63 \\
    $\mu$  & $[\si{\milli\pascal\second}]$ & 0.91 & 0.91 & 0.85 & 0.91 & 0.85 & 0.85 \\
    $Re$    & [-]        & 100  & 100 & 300 & 100 & 300 & 300 \\
    $\tau_w$ & $[\si{\pascal}]$ & 0.068 & 0.14 & 0.17 & 0.27 & 0.34 &  0.67\\
    \bottomrule
    \end{tabular}
    \caption{\textbf{Overview of the experimental parameters.} The cases are sorted by increasing $\tau_w$.}
    \label{tab:conditions}
\end{table}

\subsection{Flow conditions}
The aim of this study is to measure how the biofilm growth depends on the wall shear stress. The wall shear stress in a wide channel (without biofilm) can be estimated from plane Poiseuille flow as 
\begin{equation}
    \tau_w =\mu \frac{\partial u}{\partial z}\bigg\rvert_{z=0} = \mu \frac{6}{H} U_{b}= \mu \frac{6Q}{H^2 w},
    \label{eq:wss}
\end{equation}
where $\mu$ is the dynamic viscosity, $U_b$ the bulk velocity, $Q$ the flow rate, $H$ the channel height, and $w$ the channel width.  In our experiments, the average height of the biofilm remains below $\SI{3}{\percent}$ of the channel height. Therefore, the estimated wall-shear stress is not adjusted over time.

The wall shear stress \eqref{eq:wss} can be varied in two ways: by changing the flow rate in a given channel \citep{Kim.2013}, or by a change in channel height at a given flow rate \citep{Thomen.2017, Chun.2022}. Here, the biofilms are grown in channels of three heights ($\SI{1}{mm}, \sqrt{2}\si{mm}, \SI{2}{mm}$) and the experiments are conducted at two fixed flow rates: $\SI{1}{\milli\litre\per\second}$ and $\SI{2.63}{\milli\litre\per\second}$. The combination of multiple flow rates and channel heights results in six configurations (see Table \ref{tab:conditions}).

\begin{figure}[t]
    \centering
    \includegraphics[width=0.6\textwidth]{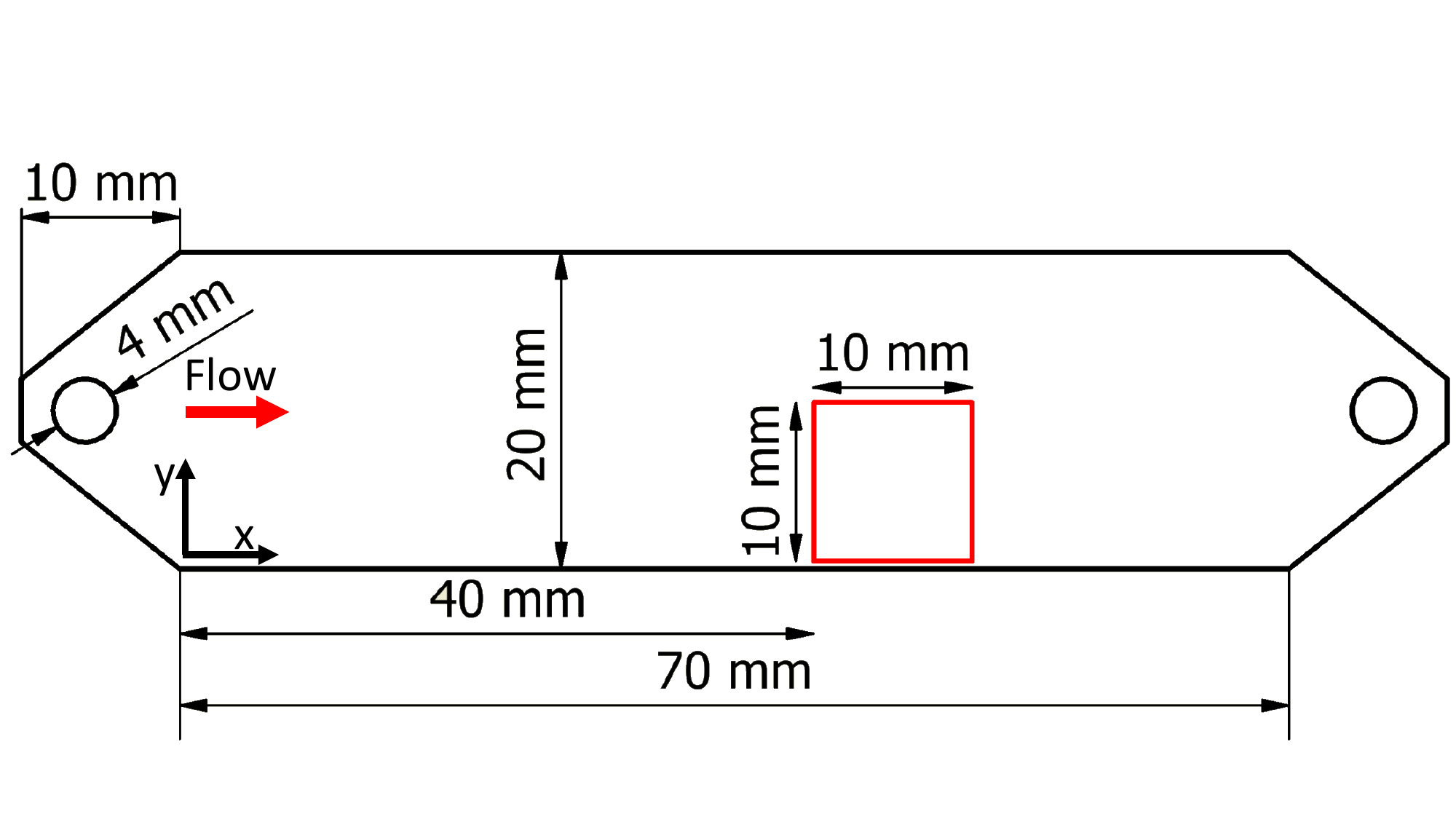}
    \caption{\textbf{Channel geometry.} The measurement area (FOV = $\SI{1}{\centi\metre\squared}$) is highlighted in red.}
    \label{fig:Channel geometry}  
\end{figure}

Figure \ref{fig:Channel geometry} shows a top view of the geometry of one channel. All channels have a width of $\SI{20}{mm}$ and a length of $\SI{70}{mm}$ with an aspect ratio (width-to-height) of at least $AR=10$, minimising the influence of the side walls. To avoid the formation of jets and thus reduce the entrance length, the medium is introduced to the channel in the vertical direction. The channels are made of milled polyoxymethylene (POM). Optical access is provided through the top of the channel, which is made of polymethyl methacrylate (PMMA). 
Due to the different loads on the pumps, the temperature inside the system stabilised at $\SI{24}{\celsius}$ and $\SI{27}{\celsius}$ respectively.
Therefore, the experiments were conducted at Reynolds numbers of $Re = {U_b D_H \rho}/{\mu}= 100$ and $Re = 300$. Here, $D_H$ is the hydraulic diameter and $\rho$ is the density of water. 
The relevant parameters for the six configurations are listed in Table \ref{tab:conditions}. 

\begin{figure}
    \centering
    \tikzset{every picture/.style={line width=0.75pt}} 
\resizebox{\textwidth}{!}{
\begin{tikzpicture}[x=0.75pt,y=0.75pt,yscale=-1,xscale=1]

\draw   (9.5,149.5) -- (139.5,149.5) -- (139.5,179.5) -- (9.5,179.5) -- cycle ;

\draw   (149.5,149.5) -- (279.5,149.5) -- (279.5,179.5) -- (149.5,179.5) -- cycle ;

\draw   (149.5,69.5) -- (279.5,69.5) -- (279.5,99.5) -- (149.5,99.5) -- cycle ;

\draw   (289.5,69.5) -- (419.5,69.5) -- (419.5,99.5) -- (289.5,99.5) -- cycle ;

\draw   (149.5,109.5) -- (279.5,109.5) -- (279.5,139.5) -- (149.5,139.5) -- cycle ;

\draw   (429.5,149.5) -- (559.5,149.5) -- (559.5,179.5) -- (429.5,179.5) -- cycle ;

\draw   (9.5,69.5) -- (139.5,69.5) -- (139.5,99.5) -- (9.5,99.5) -- cycle ;

\draw   (429.5,69.5) -- (559.5,69.5) -- (559.5,99.5) -- (429.5,99.5) -- cycle ;

\draw   (289.5,109.5) -- (419.5,109.5) -- (419.5,139.5) -- (289.5,139.5) -- cycle ;

\draw   (429.5,109.5) -- (559.5,109.5) -- (559.5,139.5) -- (429.5,139.5) -- cycle ;

\draw   (9.5,109.5) -- (139.5,109.5) -- (139.5,139.5) -- (9.5,139.5) -- cycle ;

\draw   (289.5,149.5) -- (419.5,149.5) -- (419.5,179.5) -- (289.5,179.5) -- cycle ;

\draw    (139.5,84.5) -- (149.5,84.5) ;
\draw    (139.25,165.25) -- (149.25,165.25) ;
\draw    (139,124.5) -- (149,124.5) ;
\draw    (279.5,164.65) -- (289.5,164.65) ;
\draw    (279.25,124.65) -- (289.25,124.65) ;
\draw    (279.25,84.25) -- (289.25,84.25) ;
\draw    (418.85,84.5) -- (428.85,84.5) ;
\draw    (419.1,124.75) -- (429.1,124.75) ;
\draw    (419.6,164.5) -- (429.6,164.5) ;
\draw    (560,85.5) -- (570,85.5) ;
\draw    (559.95,125.5) -- (569.95,125.5) ;
\draw    (560,165) -- (564.7,165) -- (570,165) ;
\draw    (0,85) -- (10,85) ;
\draw    (0,165) -- (10,165) ;
\draw    (0,125) -- (10,125) ;

\draw (325,155) node [anchor=north west][inner sep=0.75pt]   [align=left] {H=2mm};
\draw (45,115) node [anchor=north west][inner sep=0.75pt]   [align=left] {H=2mm};
\draw (465,115) node [anchor=north west][inner sep=0.75pt]   [align=left] {H=2mm};
\draw (325,113) node [anchor=north west][inner sep=0.75pt]   [align=left] {H=$\sqrt{2}$mm};
\draw (465,73) node [anchor=north west][inner sep=0.75pt]   [align=left] {H=$\sqrt{2}$mm};
\draw (45,73) node [anchor=north west][inner sep=0.75pt]   [align=left] {H=$\sqrt{2}$mm};
\draw (465,155) node [anchor=north west][inner sep=0.75pt]   [align=left] {H=1mm};
\draw (185,115) node [anchor=north west][inner sep=0.75pt]   [align=left] {H=1mm};
\draw (325,75) node [anchor=north west][inner sep=0.75pt]   [align=left] {H=1mm};
\draw (185,75) node [anchor=north west][inner sep=0.75pt]   [align=left] {H=2mm};
\draw (185,153) node [anchor=north west][inner sep=0.75pt]   [align=left] {H=$\sqrt{2}$mm};
\draw (45,155) node [anchor=north west][inner sep=0.75pt]   [align=left] {H=1mm};

\end{tikzpicture}
}
    \caption{The channels are ordered such that each channel height is placed in each streamwise position once.}
    \label{fig:channelGrid}
\end{figure}
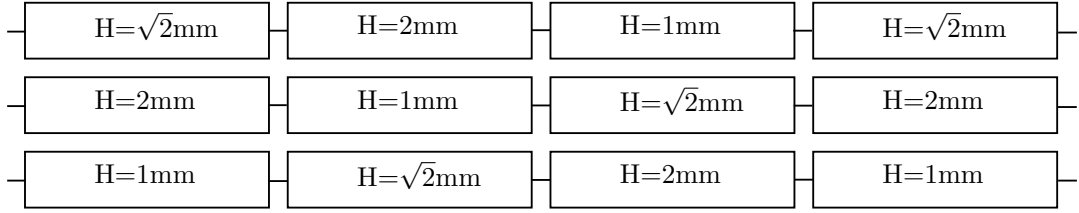

\begin{figure}
    \centering
    \tikzset{every picture/.style={line width=0.75pt}} 

\begin{tikzpicture}[x=0.75pt,y=0.75pt,yscale=-1,xscale=1]

\draw  [color={rgb, 255:red, 74; green, 144; blue, 226 }  ,draw opacity=1 ][fill={rgb, 255:red, 74; green, 144; blue, 226 }  ,fill opacity=1 ] (218.94,195.34) .. controls (218.94,198.08) and (216.72,200.3) .. (213.98,200.3) -- (185.5,200.3) .. controls (182.76,200.3) and (180.54,198.08) .. (180.54,195.34) -- (180.54,175.5) .. controls (180.54,175.5) and (180.54,175.5) .. (180.54,175.5) -- (218.94,175.5) .. controls (218.94,175.5) and (218.94,175.5) .. (218.94,175.5) -- cycle ;
\draw  [color={rgb, 255:red, 184; green, 233; blue, 134 }  ,draw opacity=1 ][fill={rgb, 255:red, 184; green, 233; blue, 134 }  ,fill opacity=1 ] (386.05,86.77) -- (431.47,90.35) -- (431.19,93.88) -- (385.78,90.31) -- cycle ;
\draw   (387,74.72) -- (432.42,78.3) -- (431.19,93.88) -- (385.78,90.31) -- cycle ;

\draw   (338.07,10.8) .. controls (338.07,6.49) and (341.56,3) .. (345.87,3) -- (369.27,3) .. controls (373.57,3) and (377.07,6.49) .. (377.07,10.8) -- (377.07,43) .. controls (377.07,43) and (377.07,43) .. (377.07,43) -- (338.07,43) .. controls (338.07,43) and (338.07,43) .. (338.07,43) -- cycle ;
\draw [color={rgb, 255:red, 255; green, 3; blue, 33 }  ,draw opacity=1 ][line width=2.25]    (357.73,43.67) -- (357.73,85.41) ;
\draw    (451,86.67) -- (432,86.17) ;
\draw    (200.6,68.8) -- (234.93,68.8) ;
\draw   (220.1,193.03) .. controls (220.1,197.49) and (216.49,201.1) .. (212.03,201.1) -- (187.81,201.1) .. controls (183.35,201.1) and (179.73,197.49) .. (179.73,193.03) -- (179.73,152.5) .. controls (179.73,152.5) and (179.73,152.5) .. (179.73,152.5) -- (220.1,152.5) .. controls (220.1,152.5) and (220.1,152.5) .. (220.1,152.5) -- cycle ;
\draw   (308.5,186.83) .. controls (308.5,173.39) and (319.39,162.5) .. (332.83,162.5) .. controls (346.27,162.5) and (357.17,173.39) .. (357.17,186.83) .. controls (357.17,200.27) and (346.27,211.17) .. (332.83,211.17) .. controls (319.39,211.17) and (308.5,200.27) .. (308.5,186.83) -- cycle ;
\draw   (320.67,174.67) .. controls (320.67,167.95) and (326.11,162.5) .. (332.83,162.5) .. controls (339.55,162.5) and (345,167.95) .. (345,174.67) .. controls (345,181.39) and (339.55,186.83) .. (332.83,186.83) .. controls (326.11,186.83) and (320.67,181.39) .. (320.67,174.67) -- cycle ;
\draw   (320.67,199) .. controls (320.67,192.28) and (326.11,186.83) .. (332.83,186.83) .. controls (339.55,186.83) and (345,192.28) .. (345,199) .. controls (345,205.72) and (339.55,211.17) .. (332.83,211.17) .. controls (326.11,211.17) and (320.67,205.72) .. (320.67,199) -- cycle ;

\draw    (357.17,187.33) -- (451,187.33) ;
\draw    (451,187.33) -- (451,107.2) -- (451,86.67) ;
\draw    (200.6,68.8) -- (200.07,152.17) ;
\draw    (220.5,186.5) -- (308.5,186.5) ;
\draw [shift={(269.5,186.5)}, rotate = 180] [fill={rgb, 255:red, 0; green, 0; blue, 0 }  ][line width=0.08]  [draw opacity=0] (8.93,-4.29) -- (0,0) -- (8.93,4.29) -- cycle    ;
\draw   (73.48,150.35) .. controls (73.49,154.77) and (69.92,158.36) .. (65.5,158.37) -- (11.5,158.48) .. controls (7.08,158.49) and (3.49,154.92) .. (3.48,150.5) -- (3.42,118.5) .. controls (3.42,118.5) and (3.42,118.5) .. (3.42,118.5) -- (73.42,118.35) .. controls (73.42,118.35) and (73.42,118.35) .. (73.42,118.35) -- cycle ;
\draw   (103.5,138.5) .. controls (103.5,124.69) and (114.69,113.5) .. (128.5,113.5) .. controls (142.31,113.5) and (153.5,124.69) .. (153.5,138.5) .. controls (153.5,152.31) and (142.31,163.5) .. (128.5,163.5) .. controls (114.69,163.5) and (103.5,152.31) .. (103.5,138.5) -- cycle ;
\draw   (107.82,146.23) .. controls (107.82,142.42) and (110.91,139.34) .. (114.72,139.34) .. controls (118.52,139.34) and (121.61,142.42) .. (121.61,146.23) .. controls (121.61,150.04) and (118.52,153.12) .. (114.72,153.12) .. controls (110.91,153.12) and (107.82,150.04) .. (107.82,146.23) -- cycle ;
\draw   (135.39,146.23) .. controls (135.39,142.42) and (138.48,139.34) .. (142.28,139.34) .. controls (146.09,139.34) and (149.18,142.42) .. (149.18,146.23) .. controls (149.18,150.04) and (146.09,153.12) .. (142.28,153.12) .. controls (138.48,153.12) and (135.39,150.04) .. (135.39,146.23) -- cycle ;
\draw   (121.95,122.11) .. controls (121.95,118.3) and (125.04,115.22) .. (128.84,115.22) .. controls (132.65,115.22) and (135.74,118.3) .. (135.74,122.11) .. controls (135.74,125.92) and (132.65,129) .. (128.84,129) .. controls (125.04,129) and (121.95,125.92) .. (121.95,122.11) -- cycle ;
\draw    (114.72,153.12) -- (142.28,153.12) ;
\draw    (134.41,118.25) -- (148.19,142.13) ;
\draw    (108.66,142.76) -- (122.44,118.88) ;

\draw    (73.5,138.5) -- (103.5,138.5) ;
\draw [shift={(93.5,138.5)}, rotate = 180] [fill={rgb, 255:red, 0; green, 0; blue, 0 }  ][line width=0.08]  [draw opacity=0] (8.93,-4.29) -- (0,0) -- (8.93,4.29) -- cycle    ;
\draw    (153.5,138.5) -- (193.5,138.5) ;
\draw    (193.5,138.5) -- (193.58,152.3) ;
\draw   (151.34,231.14) .. controls (151.34,235.56) and (147.76,239.14) .. (143.34,239.14) -- (89.34,239.14) .. controls (84.92,239.14) and (81.34,235.56) .. (81.34,231.14) -- (81.34,199.14) .. controls (81.34,199.14) and (81.34,199.14) .. (81.34,199.14) -- (151.34,199.14) .. controls (151.34,199.14) and (151.34,199.14) .. (151.34,199.14) -- cycle ;
\draw    (180.54,175.5) -- (124.54,175.5) ;
\draw [shift={(147.54,175.5)}, rotate = 360] [fill={rgb, 255:red, 0; green, 0; blue, 0 }  ][line width=0.08]  [draw opacity=0] (8.93,-4.29) -- (0,0) -- (8.93,4.29) -- cycle    ;
\draw    (124.54,175.5) -- (124.54,199.14) ;
\draw  [color={rgb, 255:red, 184; green, 233; blue, 134 }  ,draw opacity=1 ][fill={rgb, 255:red, 184; green, 233; blue, 134 }  ,fill opacity=1 ] (233.89,74.11) -- (279.3,77.68) -- (279.03,81.21) -- (233.61,77.64) -- cycle ;
\draw   (234.84,62.06) -- (280.25,65.63) -- (279.03,81.21) -- (233.61,77.64) -- cycle ;

\draw  [color={rgb, 255:red, 184; green, 233; blue, 134 }  ,draw opacity=1 ][fill={rgb, 255:red, 184; green, 233; blue, 134 }  ,fill opacity=1 ] (284.39,78.44) -- (329.8,82.01) -- (329.53,85.55) -- (284.11,81.97) -- cycle ;
\draw   (285.34,66.39) -- (330.75,69.96) -- (329.53,85.55) -- (284.11,81.97) -- cycle ;

\draw  [color={rgb, 255:red, 184; green, 233; blue, 134 }  ,draw opacity=1 ][fill={rgb, 255:red, 184; green, 233; blue, 134 }  ,fill opacity=1 ] (334.72,82.61) -- (380.14,86.18) -- (379.86,89.71) -- (334.44,86.14) -- cycle ;
\draw   (335.67,70.56) -- (381.09,74.13) -- (379.86,89.71) -- (334.44,86.14) -- cycle ;

\draw    (279.34,73.14) -- (284.9,73.95) ;
\draw    (330.09,77.89) -- (335.65,78.7) ;
\draw    (380.59,82.14) -- (386.15,82.95) ;

\draw (93.34,209.58) node [anchor=north west][inner sep=0.75pt]   [align=left] {Waste};
\draw (9.34,126.98) node [anchor=north west][inner sep=0.75pt]   [align=left] {Medium};
\draw (89.5,90) node [anchor=north west][inner sep=0.75pt]   [align=left] {Feed pump};
\draw (289.67,210.33) node [anchor=north west][inner sep=0.75pt]   [align=left] {Gear pump};
\draw (340,10.8) node [anchor=north west][inner sep=0.75pt]   [align=left] {OCT};
\draw (164.83,209.83) node [anchor=north west][inner sep=0.75pt]   [align=left] {Reservoir};
\draw (295.33,95) node [anchor=north west][inner sep=0.75pt]   [align=left] {Flow cells};

\end{tikzpicture}
    \caption{\textbf{Setup of one flow loop.} Three independent flow loops are used.}
    \label{fig:sketch}
\end{figure}
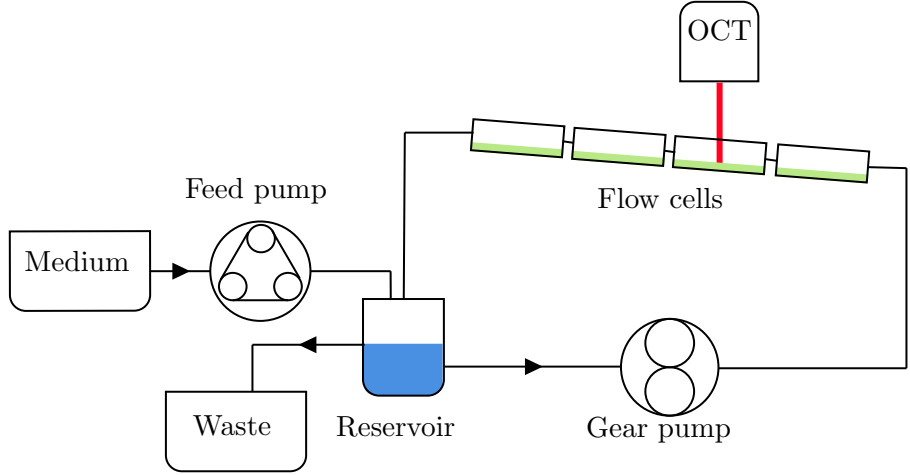

Twelve channels are set up in three independent flow loops, where each loop consists of four channels in series. The channels are ordered such that each channel height is placed once in each streamwise position. The position of the channels within the larger setup is visualised in Figure \ref{fig:channelGrid}.  
One flow loop is shown in detail in Figure \ref{fig:sketch}. Each flow loop is driven by a magnetic gear pump (Niemzik PAT, Haan, Germany). The cultivation medium is recirculated and fresh medium is extracted from a $\SI{10}{\litre}$ bottle by a peristaltic pump and added to the reservoir. The medium in the reservoir is constantly aerated to ensure sufficient oxygen supply. Mixing is provided by magnetic stirrers. The volume in the reservoirs is kept constant by an outlet at the target height. The cumulative residence time of the medium in four flow cells is below $\SI{15}{\second}$.
Given these conditions, 
we can assume a steady state in the bulk liquid without nutrient depletion causing a concentration gradient in the streamwise direction.

During the experiment, the flow rate in each flow loop is monitored periodically. Once the flow rate drops sharply, indicating rapid blockage somewhere in the flow loop, the measurement is stopped. This occurred on the eighth day. Additionally, measurements are inspected for artefacts, such as bubbles or large aggregates of biofilm getting stuck in the flow cell. Scans including these are removed from the evaluation.

\subsection{Culture preparation}
\textit{Bacillus subtilis} is taken from frozen stock, plated, and a single colony is resuspended in liquid LB (Lennox) medium. The liquid culture is incubated overnight at $\SI{30}{\celsius}$ with shaking at $\SI{140}{rpm}$ and diluted to an optical density of $OD_{600} = 0.1$. The working liquid in the channel is a volumetric 1:100 solution of LB (Lennox) broth in tap water (for mineral content, see \cite{KarlsruheWater2023}, data for July 2023). The chemical oxygen demand (COD) of the diluted medium is approximately $\SI{250}{\milli\gram\per\liter}$. The inoculum is introduced to the flow cells at the target flow rate $Q$. After one hour, medium replacement is started at a flow rate of $Q_r = \SI{1}{\milli\liter\per\minute}$ corresponding to a replacement rate of five times per day.
The temperature in the medium reservoir is monitored periodically.

\subsection{OCT measurements}

The biofilm structure is scanned without interruption of the flow, i.e. without deforming the biofilm due to changing conditions. Measurements are acquired using an OCT system (Thorlabs Ganymede GAN610-SP5) with a central wavelength $\SI{930}{\nano\meter}$ and an LSM04 objective lens. The OCT head is moved to predetermined measurement positions, $\SI{45}{mm}$ downstream from the inlet, by the automated traverse system introduced by Gierl et al. \citep{Gierl.2020}. The traverse system is an evoBot with a positional accuracy of $\SI{0.1}{\milli\meter}$ \citep{Nejatimoharrami.2017}. Measurements are taken with a field of view of $\SI{10}{mm} \times \SI{10}{mm}$ (streamwise $\times$ cross-stream)  and a lateral pixel size of $\SI{12}{\micro\meter}$. The sample rate of the OCT is set to $\SI{100}{kHz}$. The axial pixel size of the OCT in water is $\SI{2.1}{\micro\meter}$. A-scan averaging is set to two. Flat interfaces perpendicular to the OCT beam, such as the air-PMMA interface and the PMMA-medium interface, can cause strong reflections and autocorrelation noise in the OCT signal. Therefore, the channels are mounted on a tilted base-plate. The angle is set to $\SI{5}{\degree}$. The first measurements are taken two hours after inoculation and then every twelve hours. The twelve channels are scanned sequentially within one hour.

\subsection{Data processing and thresholding algorithm} \label{sec:dataproc}
\begin{figure}
    \begin{subfigure}{0.32\textwidth}
        \frame{\includegraphics[width=\textwidth,trim={0 0 0 2.5cm},clip]{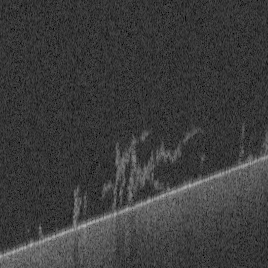}}
        \subcaption{Raw image}
        \label{fig:rawImage}
    \end{subfigure}
    \begin{subfigure}{0.32\textwidth}
        \frame{\includegraphics[width=\textwidth,trim={0 0 0 2.5cm},clip]{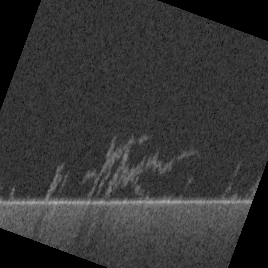}}
        \subcaption{Rotated image}
        \label{fig:rotatedImage}
    \end{subfigure}
    \begin{subfigure}{0.32\textwidth}
        \frame{\includegraphics[width=\textwidth,trim={0 0 0 2.5cm},clip]{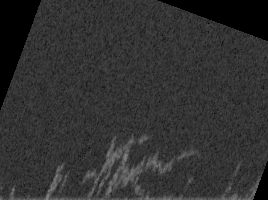}}
        \subcaption{Substratum detected}
        \label{fig:substratumDetected}
    \end{subfigure}\\
    
    \begin{subfigure}{0.32\textwidth}
        \frame{\includegraphics[width=\textwidth,trim={0 0 0 2.5cm},clip]{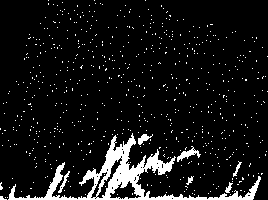}}
        \subcaption{Thresholded image}
        \label{fig:thresholdedImage}
    \end{subfigure}
    \begin{subfigure}{0.32\textwidth}
        \frame{\includegraphics[width=\textwidth,trim={0 0 0 2.5cm},clip]{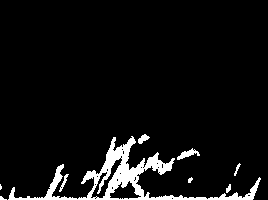}}
        \subcaption{Noise removed}
        \label{fig:noiseRemoved}
    \end{subfigure}
    \begin{subfigure}{0.32\textwidth}
        \frame{\includegraphics[width=\textwidth]{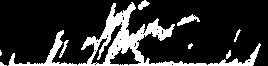}}
        \subcaption{Final image}
        \label{fig:finalImage}
    \end{subfigure}
\caption{\textbf{Processing of a sample scan.} The raw scan is rotated and the substratum is aligned with the bottom of the scan. Then, the image is binarised, denoised, and trimmed to the extent of the biofilm. These images are not to scale to improve visibility.}
\label{fig:imageProcessing}
\end{figure}

The OCT imaging software ThorImage returns files in .oct format. This format is an archive that contains metadata and the raw image files in 32-bit floating point greyscale. The raw images are extracted automatically and prepared for processing in ImageJ \citep{Schindelin.2012}. The processing steps are visualised in Figure \ref{fig:imageProcessing}. Since the channels are placed at an angle, the image stacks have to be rotated for the substratum to be oriented horizontally (Figure \ref{fig:rawImage}). ImageJ is chosen for this step because of its speed at rotating stacks of images. First, the images are converted to 8-bit greyscale, then they are rotated by preset angles along two axes (Figure \ref{fig:rotatedImage}). The resulting image stacks are saved.

Further processing is performed using custom Python scripts, which are available at \url{https://www.bagherigroup.com/research/open-source-codes/}.
First, the substratum is detected by finding the point of highest intensity along the vertical axis and then applying a generous median filter with a radius of eleven pixels. The optomechanical setup of the OCT system causes minor warping in the planar substratum that cannot be removed by the predefined calibration function. Here, this remaining warping is removed by aligning the detected substratum at the bottom of the scan (Figure \ref{fig:substratumDetected}). 

We developed a thresholding algorithm for our OCT data since existing schemes such as Otsu's method struggle to accurately detect the biofilm. OCT scans of empty volumes are noisy, with the signal following a Gaussian distribution. The signal that represents biofilm has an intensity that is larger than the mode of the intensity distribution. 
When the amount of biofilm within a scan is too small to create a bimodal distribution, there is an inflection point marking the beginning of the biofilm data. This fact is used in the following algorithm. First, the mode of the distribution is detected. Then, the maximum of the second derivative of the intensity distribution to the right of the mode is marked as a candidate threshold. 
A further offset of three intensity levels is required to reliably match the results of manual thresholding. The resulting threshold is used to binarise the image stack (Figure \ref{fig:thresholdedImage}). Afterwards, salt-and-pepper noise is removed using a filter similar to the \textit{Remove Outliers} function in ImageJ (Figure \ref{fig:noiseRemoved}). This filter acts as a median filter removing small biofilm structures, without filling in any gaps. Finally, the image is trimmed from the top to the section containing biofilm to minimise file sizes (Figure \ref{fig:finalImage}). Once all scans have been processed, consecutive scans of the same flow cell are aligned using cross-correlation-based image registration.

\subsection{Measures of biofilm morphology}
\begin{figure}
    \centering
    \tikzset{every picture/.style={line width=0.75pt}} 

\begin{tikzpicture}[x=0.75pt,y=0.75pt,yscale=-1,xscale=1]

\draw    (20.25,159.33) -- (218.36,159.33) ;
\draw [shift={(221.36,159.33)}, rotate = 180] [fill={rgb, 255:red, 0; green, 0; blue, 0 }  ][line width=0.08]  [draw opacity=0] (10.72,-5.15) -- (0,0) -- (10.72,5.15) -- (7.12,0) -- cycle    ;
\draw    (20,159.33) -- (20,49.3) ;
\draw [shift={(20,46.3)}, rotate = 90] [fill={rgb, 255:red, 0; green, 0; blue, 0 }  ][line width=0.08]  [draw opacity=0] (10.72,-5.15) -- (0,0) -- (10.72,5.15) -- (7.12,0) -- cycle    ;
\draw  [fill={rgb, 255:red, 184; green, 233; blue, 134 }  ,fill opacity=1 ] (108.53,68.93) .. controls (128,56) and (166.73,62.87) .. (140.2,85.6) .. controls (113.67,108.33) and (109.77,124.63) .. (99,139.33) .. controls (88.23,154.03) and (85.86,159.29) .. (69,159.33) .. controls (52.14,159.38) and (78.53,159.6) .. (39,159.33) .. controls (-0.53,159.07) and (89.07,81.87) .. (108.53,68.93) -- cycle ;
\draw    (121.36,66.92) -- (121.36,72) -- (121.36,100.25) ;
\draw [shift={(121.36,103.25)}, rotate = 270] [fill={rgb, 255:red, 0; green, 0; blue, 0 }  ][line width=0.08]  [draw opacity=0] (5.36,-2.57) -- (0,0) -- (5.36,2.57) -- cycle    ;
\draw [shift={(121.36,63.92)}, rotate = 90] [fill={rgb, 255:red, 0; green, 0; blue, 0 }  ][line width=0.08]  [draw opacity=0] (5.36,-2.57) -- (0,0) -- (5.36,2.57) -- cycle    ;
\draw    (131.86,65.42) -- (131.86,156.29) ;
\draw [shift={(131.86,159.29)}, rotate = 270] [fill={rgb, 255:red, 0; green, 0; blue, 0 }  ][line width=0.08]  [draw opacity=0] (5.36,-2.57) -- (0,0) -- (5.36,2.57) -- cycle    ;
\draw [shift={(131.86,62.42)}, rotate = 90] [fill={rgb, 255:red, 0; green, 0; blue, 0 }  ][line width=0.08]  [draw opacity=0] (5.36,-2.57) -- (0,0) -- (5.36,2.57) -- cycle    ;
\draw    (31.13,162.5) -- (78.38,162.5) ;
\draw [shift={(78.38,162.5)}, rotate = 180] [color={rgb, 255:red, 0; green, 0; blue, 0 }  ][line width=0.75]    (0,2.24) -- (0,-2.24)   ;
\draw [shift={(31.13,162.5)}, rotate = 180] [color={rgb, 255:red, 0; green, 0; blue, 0 }  ][line width=0.75]    (0,2.24) -- (0,-2.24)   ;
\draw  [fill={rgb, 255:red, 184; green, 233; blue, 134 }  ,fill opacity=1 ] (170.63,66) .. controls (176.94,65.49) and (182,69) .. (203.13,68) .. controls (224.25,67) and (260,69) .. (226.38,75.75) .. controls (192.75,82.5) and (196.13,74.5) .. (174.13,75.75) .. controls (152.13,77) and (147.38,80.75) .. (148.32,73.83) .. controls (149.27,66.9) and (148.13,70.47) .. (148.13,67.47) .. controls (148.13,64.47) and (164.31,66.51) .. (170.63,66) -- cycle ;
\draw    (190,170) -- (200,160) ;
\draw    (170,169.93) -- (180,159.93) ;
\draw    (150,169.93) -- (160,159.93) ;
\draw    (130,169.93) -- (140,159.93) ;
\draw    (110,169.87) -- (120,159.87) ;
\draw    (90,169.87) -- (100,159.87) ;
\draw    (180.38,155.5) -- (180.38,148.5) ;

\draw (209.75,165) node [anchor=north west][inner sep=0.75pt]   [align=left] {x};
\draw (5,48) node [anchor=north west][inner sep=0.75pt]   [align=left] {z};
\draw (104.75,78) node [anchor=north west][inner sep=0.75pt]   [align=left] {$\displaystyle T$};
\draw (132.25,100) node [anchor=north west][inner sep=0.75pt]   [align=left] {$\displaystyle h$};
\draw (57.5,120) node [anchor=north west][inner sep=0.75pt]   [align=left] {base};
\draw (24.75,165) node [anchor=north west][inner sep=0.75pt]   [align=left] {footprint};
\draw (165,80) node [anchor=north west][inner sep=0.75pt]   [align=left] {streamer};
\draw (141,133) node [anchor=north west][inner sep=0.75pt]   [align=left] {substratum};

\end{tikzpicture}
    \caption{\textbf{Different measures of the vertical extent change the influence of voids.} The biofilm height $h$ includes empty spaces, the biofilm thickness $T$ omits voids.}
    \label{fig:heightThickness}
\end{figure}
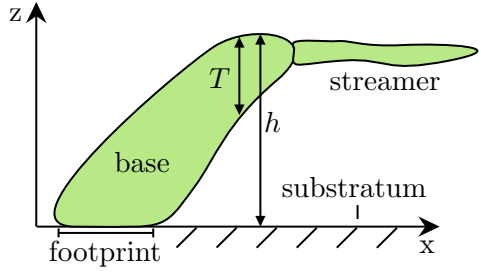

In this section, we define several measures to characterize how the spatial distribution of biofilm changes in time. 
Let $I(x,y,z,t)$ denote the intensity of the OCT signal with the position of the substratum defined as $z=0$. Biofilm exists at a position $(x,y)$ if an OCT signal at these coordinates exceeds a threshold that is determined using the algorithm described in Section \ref{sec:dataproc}. 
The spatial distribution of biofilm $b(x,y,t)$ is given by
\begin{equation}
    b(x,y,z,t) = \begin{cases}
    1,& \text{if } I\geq I_{thresh}\\
    0,              & \text{otherwise}
\end{cases}
\end{equation}
The vertical extent of the biofilm can be described using the two different measures defined in Figure \ref{fig:heightThickness}. The biofilm height $h(x,y,t)$ describes the height of the bulk-biofilm interface above the substratum. It is defined as the distance from the substratum to the highest point that contains biofilm.
The biofilm thickness $T$ is defined as the height of the biofilm excluding any voids, characterising the amount of biofilm at each position. It is given by
\begin{equation}
    T(x,y,t) = \int_0^{h} b(x,y,z,t)\, \mathrm{d}z.
    \label{eq:thickness}
\end{equation}
The biofilm volume $V_b$ inside a measurement area is given by 
\begin{equation}
    V_b(t) = \int_{V} b(x,y,z,t)\, \mathrm{d}V.
    \label{eq:biovolume}
\end{equation}
This measure only considers voxels that contain biofilm and thus accurately discards voids within and beneath the biofilm. The biofilm volume can be normalised using the area of the field of view $A$ (red box in Figure \ref{fig:Channel geometry}), yielding an equivalent mean biofilm thickness
\begin{equation}
    \overline{T}(t) = \frac{V_b(t)}{A}. 
    \label{eq:normalized_thickness}
\end{equation}
Finally, the substratum coverage, which describes the proportion of the substratum that is directly covered by biofilm, is defined as 
\begin{equation}
    SC(t) = \frac{1}{A}\int_{A} b(x,y,0,t)\, \mathrm{d}A.
    \label{eq:SC}
\end{equation}
In the following, the quantities $h(x,y)$, $T(x,y)$, $V_b$, $\overline{T}(x,y)$ and $SC$ are used to characterise different aspects of the biofilm morphology.

\section{Results}
\subsection{Biofilm morphology}\label{sec:morphology}
We begin by investigating how the morphology of the biofilm changes over time for different values of wall shear stress. 

\begin{figure}

\begin{subfigure}{0.9\textwidth}
    \centering
    \includegraphics[width=\textwidth]{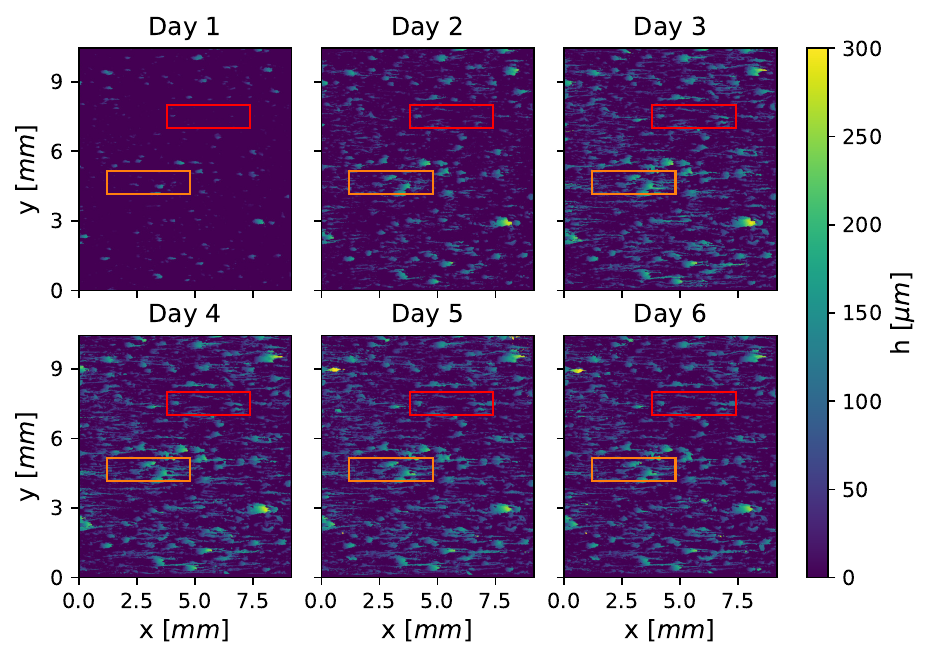}
    \caption{\textbf{Changes in biofilm height over one week.} The individual biofilm structures grow over time, as new ones appear.}
    \label{fig:timeSeries}
\end{subfigure}
\\
\begin{subfigure}{0.9\textwidth}
    \centering  
    \input{figs/streamerDevelopment}
    \caption{\textbf{Streamer development.} Region 1 corresponds to the red rectangle, and region 2 corresponds to the orange rectangle in Figure \ref{fig:timeSeries}. The boxes highlight the development of one streamer each.}
    \label{fig:streamerDevelopment}
\end{subfigure}
\caption{Biofilm growth and streamer development. The flow direction is from left to right.}
\end{figure}

\subsubsection*{Development of a sample biofilm in time}
The development of a sample biofilm from case 1 (see Tab. \ref{tab:conditions}) over six days is shown in Figure \ref{fig:timeSeries}. The top-left image shows the biofilm after twelve hours, followed by a time step of $\SI{24}{\hour}$ between each consecutive image.
During the first day, simple structures with a height of up to $\SI{175}{\micro\meter}$ form in a sparse pattern. 
Over time, the structures grow, increasing in size and complexity. 
Starting from the second day, the typical structure becomes wider and may begin developing a streamer, i.e. a thin filament that extends downstream from its tip. 
Two regions that contain significant streamer development are shown in detail in Figure \ref{fig:streamerDevelopment}. These regions correspond to the highlighted boxes in Figure \ref{fig:timeSeries}.

The biofilm morphology can qualitatively be described as a collection of microcolonies, schematically shown in Figure \ref{fig:heightThickness}. Each microcolony consists of one base -- shaped as a leaning pillar -- and may have a streamer attached to the tip of the base. 
Our observations indicate that the base and streamer grow on different time scales. 
As shown in Figure \ref{fig:streamerDevelopment}, the base develops first on a relatively slow time scale (days). Once the base has grown sufficiently large, a streamer develops relatively fast (hours), increasing in length significantly over the next $24-48 \si{\hour}$.
These streamers are aligned with the flow, with adjacent structures creating the appearance of continuous lines in the streamwise direction. This is clearly shown in region 1 (left frame in Figure \ref{fig:streamerDevelopment}), where streamers originate from three separate biofilm structures that are connected into one aggregate.
During days 4 to 6, the number of microcolonies increases over time, progressively filling the measurement area. No major sloughing events are observed. One time series of height maps from each case is available in the appendix.

\begin{figure}
\begin{subfigure}{\textwidth}
\centering
    \includegraphics[width=\textwidth]{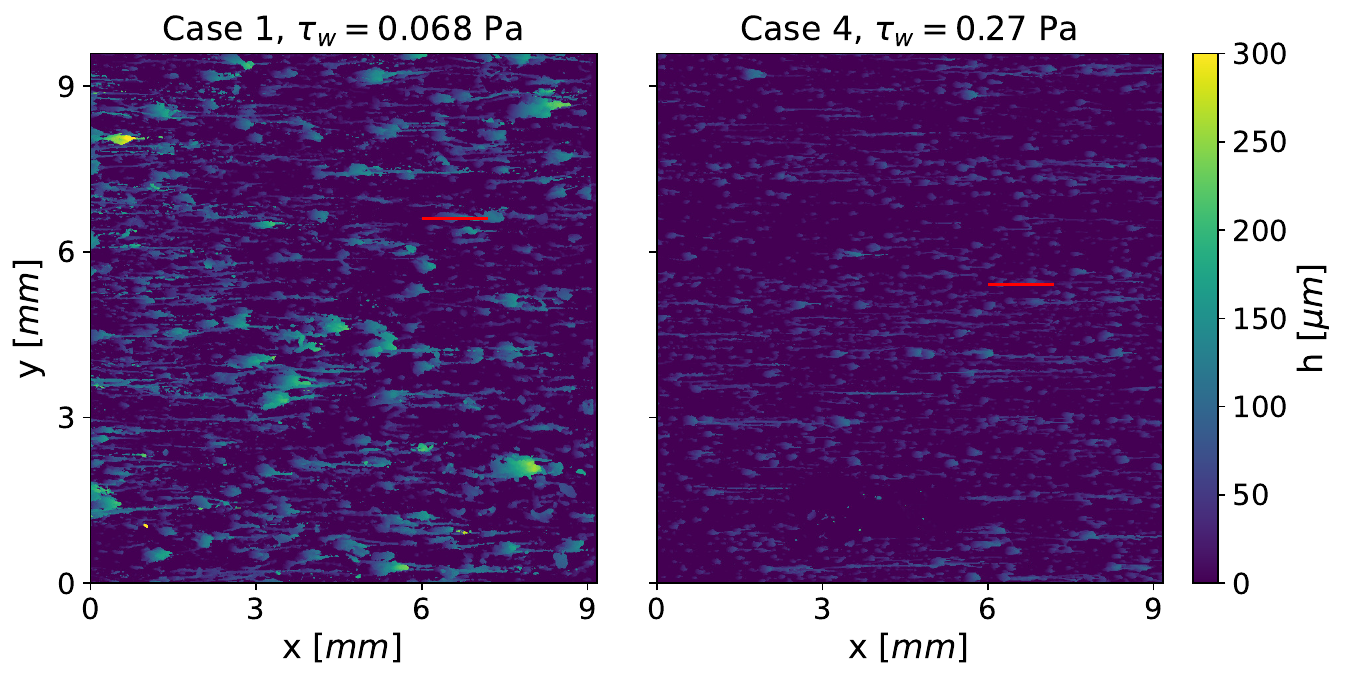}
    \caption{\textbf{Height maps of two sample biofilms after 6 days.} The flow direction is oriented from left to right.}
    \label{fig:hMaps}
\end{subfigure}   

\begin{subfigure}{\textwidth}
\centering
    \includegraphics[width=\textwidth]{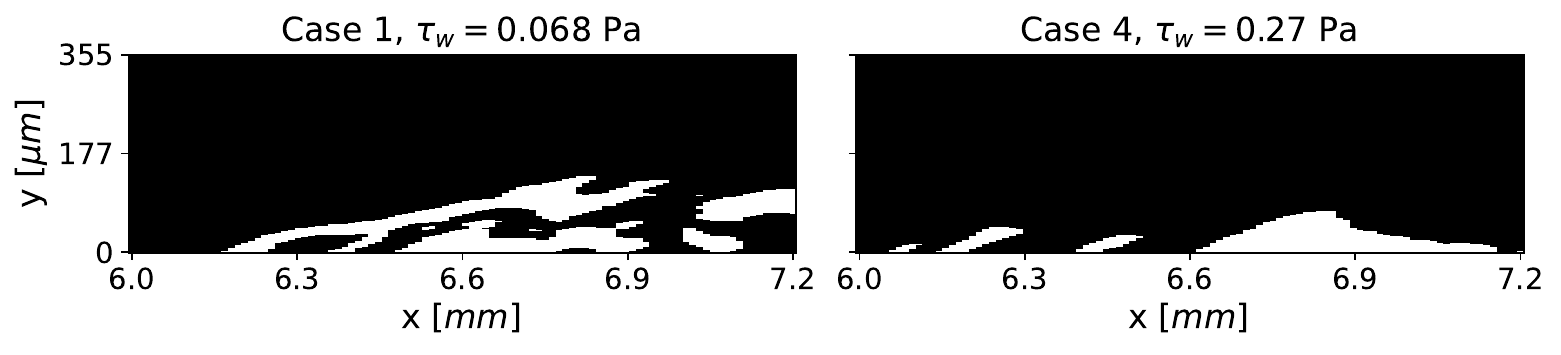}
    \caption{\textbf{Representative structures corresponding to red lines in Figure \ref{fig:hMaps}}. Under high wall shear stress, the biofilm forms more compact structures.}
    \label{fig:typicalStructures}
\end{subfigure}
\caption{Comparison of two biofilms grown under different hydrodynamic conditions: Case 1 \& 4.}
\label{fig:twoBiofilms}
\end{figure}

Streamers are known to form at sharp edges and extend downstream, specifically in positions with converging secondary flows and high shear stress due to the bulk flow \citep{Rusconi.2010, Drescher.2013}. Here, the streamers originate on the leeward side, close to the tip of the base structures. At this location, the flow around the typical base structure, which can be approximated by a leaning pillar, is likely showing such features.
It has been suggested that the expansion of streamers is driven by the attachment of planktonic cells, rather than by growth of the cells within the streamer \citep{Drescher.2013}. This rapid extension is also observed here, see Figure \ref{fig:streamerDevelopment}. No streamer motion is observed, which is consistent with the low-velocity measurements conducted in \citep{Stoodley.1998}.
For Reynolds numbers below $Re < 100$, based on the diameter of the base structure to which the streamer is attached, the drag force experienced by the streamer increases with its length \citep{Taherzadeh.2010}. Here, this Reynolds number lies below $Re < 10$. The streamers are thus likely limited in length, with the increasing drag force preventing further expansion. 

\subsubsection*{Influence of wall shear on biofilm morphology}
To investigate the influence of the wall shear stress on the biofilm morphology, we compare the structure of two biofilms. These biofilms, shown in Figure \ref{fig:hMaps}, are six days old and have grown under different conditions (cases 4 and 1 in Table \ref{tab:conditions}). The biofilm that has been grown under a low wall shear stress of $\tau_w = \SI{0.068}{\pascal}$ formed structures that are both taller and wider than the structures of biofilm grown under $\tau_w = \SI{0.27}{\pascal}$. Additionally, the streamers formed in case 1 are much larger, both in width and length, than the streamers formed in case 4. A side view of the structures that are marked by red lines in Figure \ref{fig:hMaps} is shown in Figure \ref{fig:typicalStructures}. These structures are representative of common structures in the corresponding case. Most structures are leaning in the direction of the flow, with the size of the structure decreasing with increasing wall shear stress. While most of the structures in case 4 consist of small, leaning pillars, the structures in case 1 are much more complex. Large sheets of biofilm extend from small contact points with the substratum. The segment of biofilm in case 1 that appears to be unconnected to the substratum is connected to the remaining biofilm on a neighbouring slice. The formation of more compact biofilms under increased wall shear stress is consistent with previous observations \citep{Simoes.2007, Wagner.2010}.
\begin{figure}

    \includegraphics[width=\textwidth]{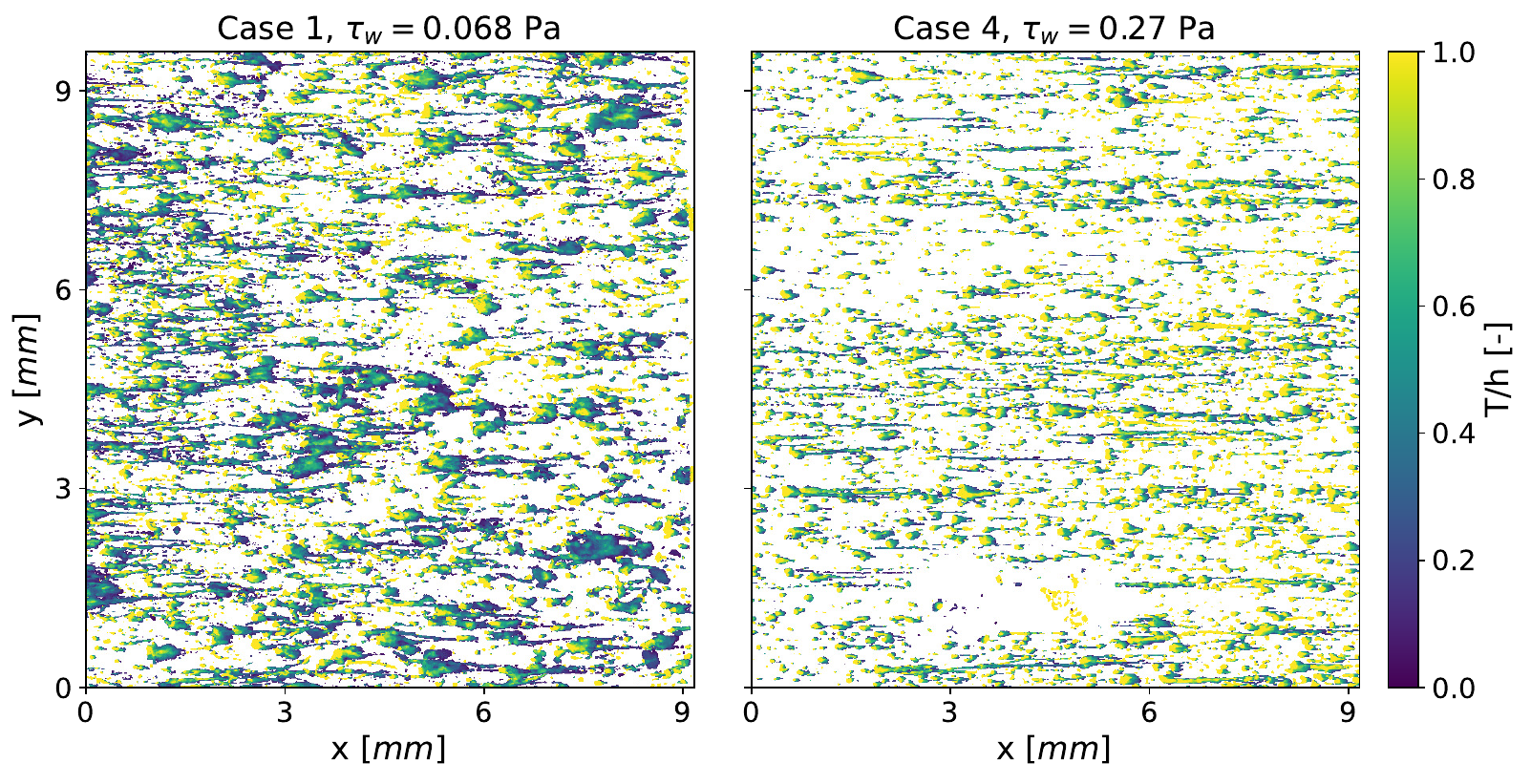}
    \caption{\textbf{Material distribution of two sample biofilms after 6 days.} Low values indicate proportionally large voids within or beneath the biofilm.}
    \label{fig:thickOverH}
\end{figure}

To quantify this observation, 
the vertical distribution of the biofilm, measured by the biofilm thickness over the biofilm height ${T}/{h}$, is shown in Figure \ref{fig:thickOverH}. Here, a value of one (yellow) represents an uninterrupted column of biofilm from the substratum to the liquid-biofilm interface, whereas a value near zero (dark green) corresponds to a thin film with a void beneath, such as in a streamer.
The microcolony structure of Figure \ref{fig:heightThickness} is particularly evident at the higher wall-shear stress (case 4 in Figure \ref{fig:thickOverH}, right frame). The bright yellow regions have the shape of a pillar, corresponding to the base structures that are fully attached to the substratum and tilted in the streamwise direction.
At the tip and downstream of many of the pillars, we observe long filamentous threads (dark green regions) in the streamwise direction, which indicates streamers that are disconnected from the substratum. 
In case 1 (Figure \ref{fig:thickOverH}, left frame), some of the microcolonies are characterised by large voids beneath the biofilms (large dark green regions) and appear as thin sheets.
Most microcolonies are attached to the substratum only in a small region on their upstream side (visible as scattered point-like yellow regions). The typical microcolony structure consisting of tilted pillars with an attached streamer represents a consistent trend throughout all the datasets. 

\begin{figure}
    \begin{subfigure}[t]{0.49\textwidth}
        \includegraphics[width=\textwidth]{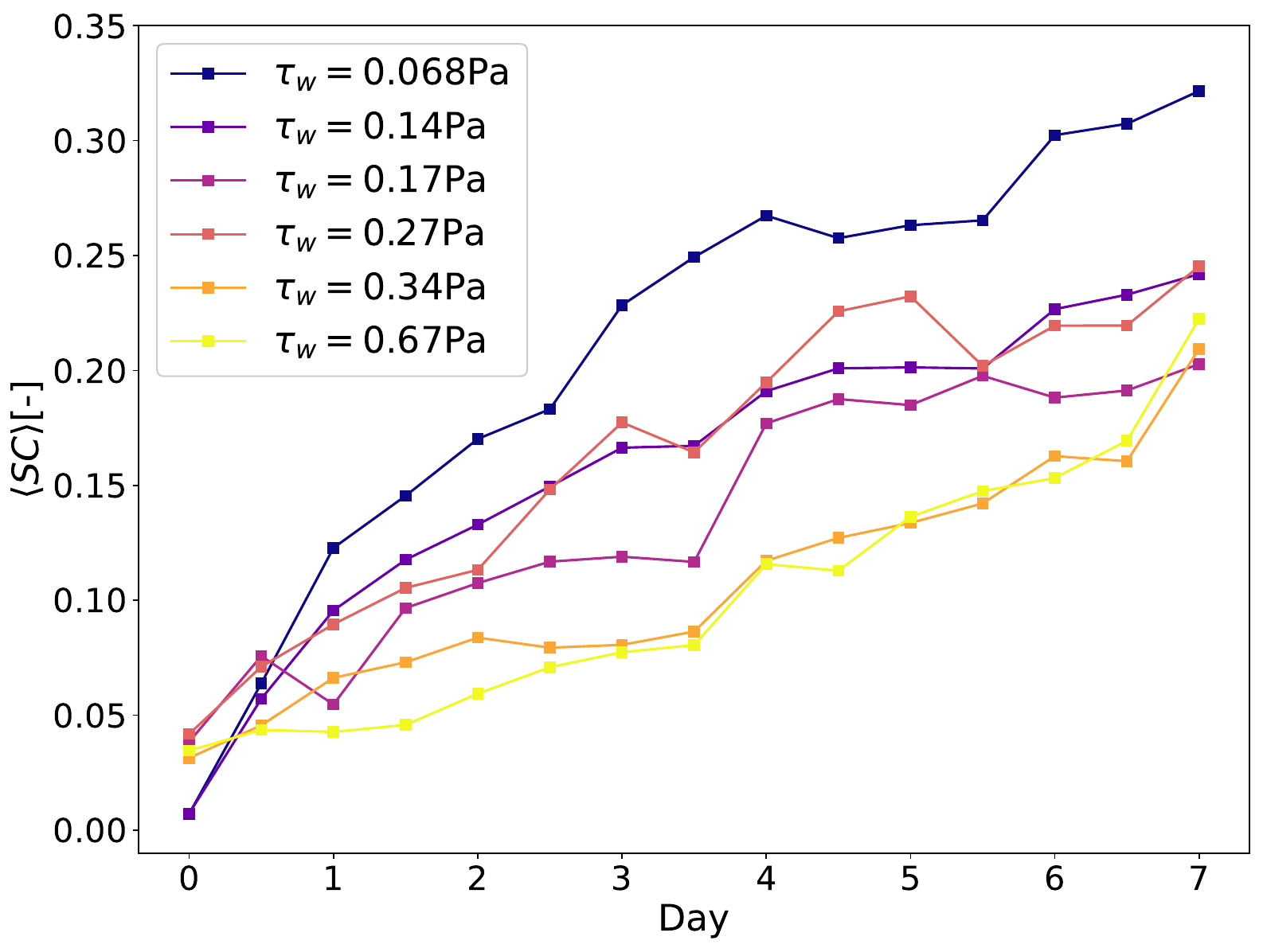}
        \caption{\textbf{Coverage over time}. The values correspond to the mean value of all replicates.} 
        \label{fig:coverageUnscaled}
    \end{subfigure}
    \begin{subfigure}[t]{0.49\textwidth}
        \includegraphics[width=\textwidth]{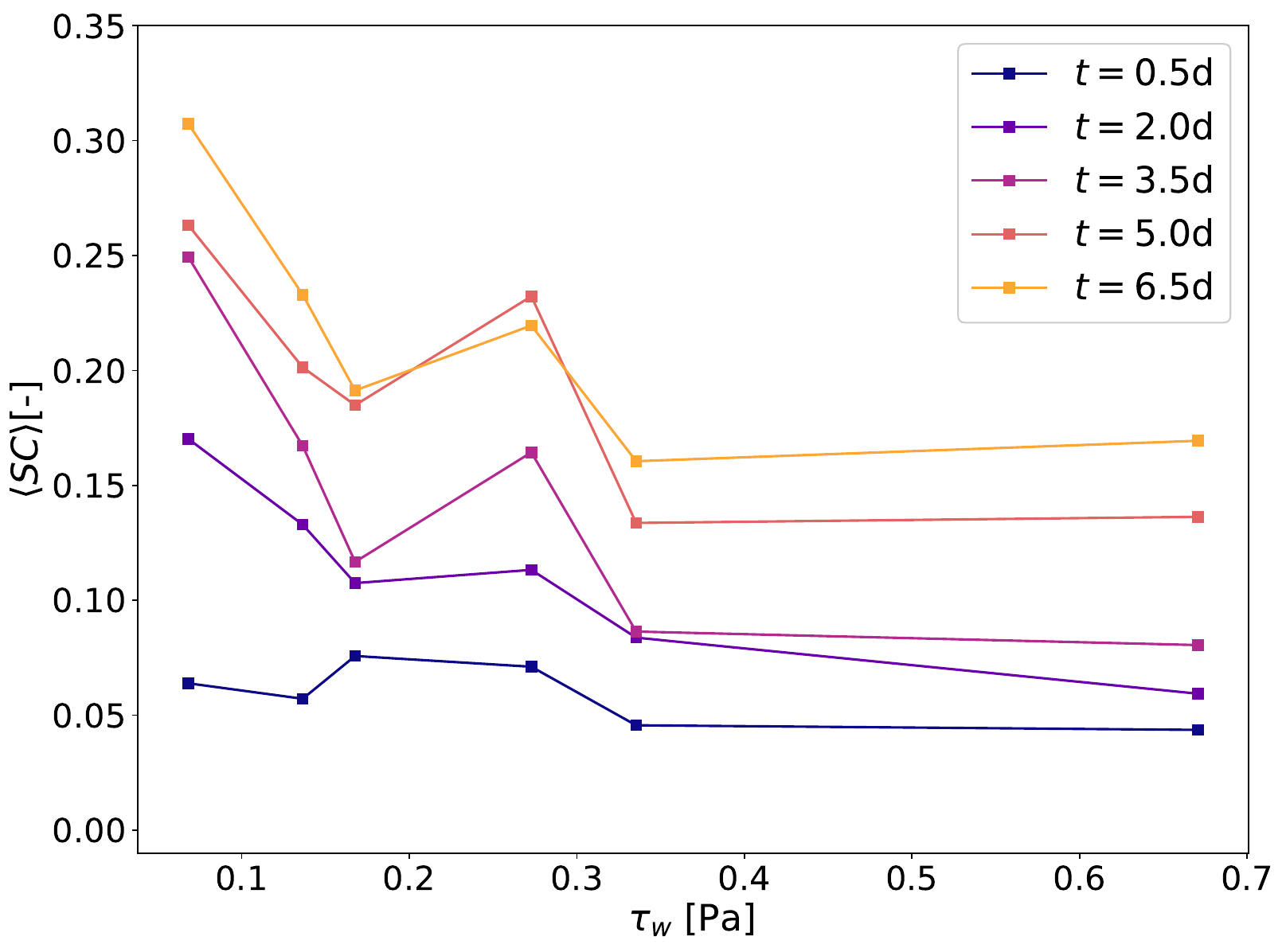}
        \caption{\textbf{Coverage over wall shear stress}. The values correspond to the mean value of all replicates. Data is shown for five regularly spaced measurements.}
        \label{fig:coverageOverTau}
    \end{subfigure}
    \caption{Change in substratum coverage}
    \label{fig:coverageData}
\end{figure}

\subsubsection*{Changes in the substratum coverage}
The observation that the coverage of the substratum decreases with increasing shear stress suggests a connection between the fluid forces acting on the biofilm and the attachment force fixing the biofilm onto the substratum. In low shear rates, a small contact patch is able to support large biofilm structures, whereas high shear rates limit the size and shape of the structures.

\begin{figure}
    \centering
    \includegraphics[width=0.5\textwidth]{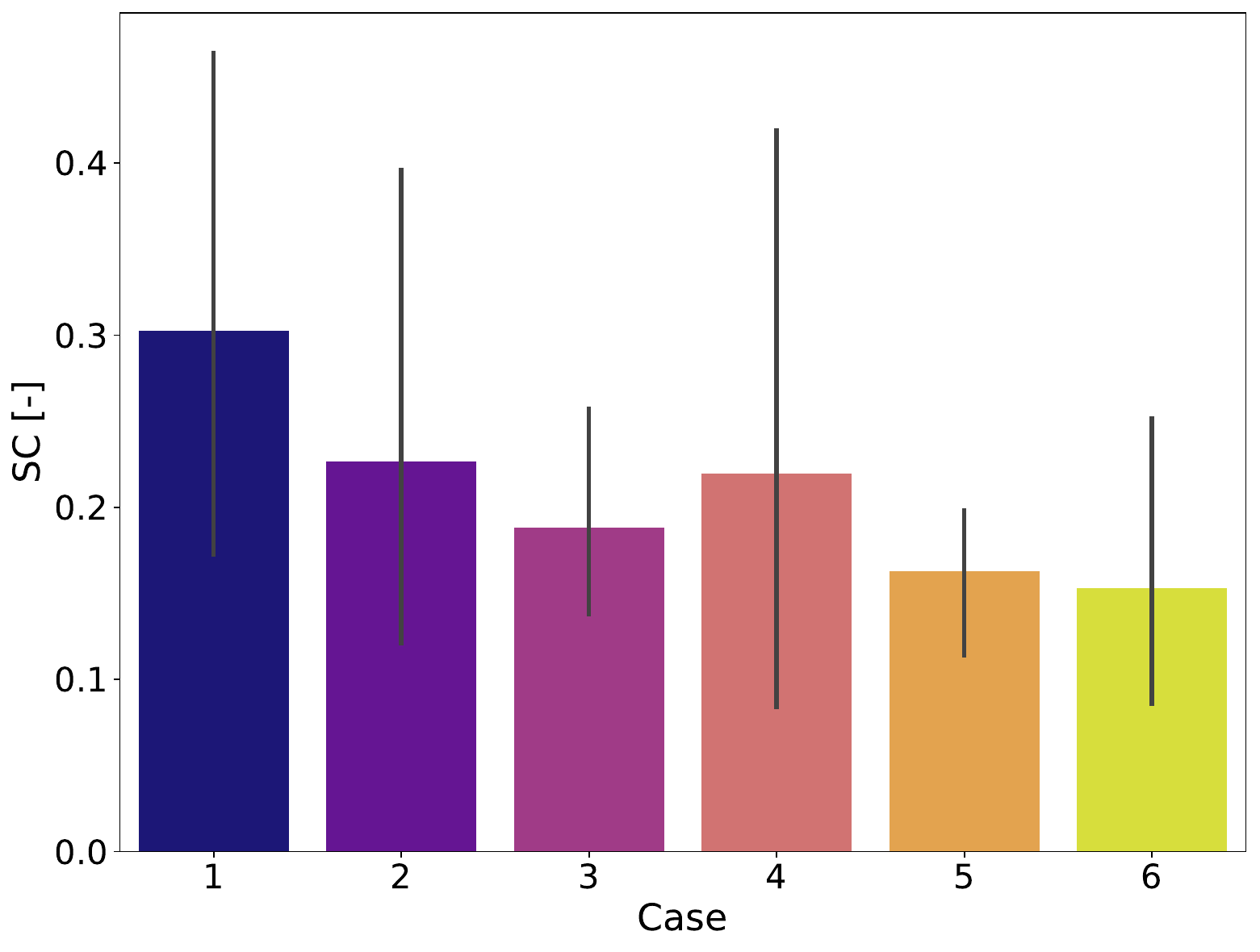}
    \caption{\textbf{Substratum coverage on day six.} The error bars represent the $\SI{95}{\percent}$ confidence interval.}
    \label{fig:reproducibility-cov}
\end{figure}

The change in substratum coverage with time for the six configurations is shown in Figure \ref{fig:coverageData}. The reported data represent the ensemble average, i.e. values obtained by averaging $SC$ of all replicates.
During the first seven days, the substratum coverage increases essentially monotonically. 
On day 4 however, the substratum coverage of the lowest-shear biofilms decreases; the same occurs on day 5 for the biofilms grown under $\tau_w = \SI{0.34}{\pascal}$. The reduction in substratum coverage is likely due to sloughing events, where patches of biofilm are detached from the substratum. These sloughing events are typical for biofilms and are often a source of variation between replicates of the same set of conditions \citep{Lewandowski.2004}. 
Our results are consistent with previous measurements by Gierl et al. who studied the growth of \textit{Bacillus subtilis} at a wall shear stress similar to case 1 \citep{Gierl.2020}. There, the surface coverage reached even higher values closing in to $SC \approx 1$. 
However, Gierl et al. defined $SC$ differently than Equation \eqref{eq:SC} as they used the maximum intensity projection of the biofilm along the vertical axis instead of the biofilm directly in contact with the substratum.
The spread of the substratum coverage on day 6 is shown in Figure \ref{fig:reproducibility-cov}, where the $\SI{95}{\percent}$ confidence intervals are reported. 
It is readily apparent that with increasing wall shear stress the substratum coverage decreases, which is consistent with previous findings \citep{Chun.2022,Chang.2020}. 
The increase in substratum coverage in case 4 is caused by one outlier, with a coverage that is 3.5 times as large as the mean of the other measurements. Note also that the increment in shear stress going from case 1 to 6 is not constant.

Figure \ref{fig:coverageOverTau} shows $SC$ as a function of the wall shear stress at different times. 
We observe that the substratum coverage in the first 12h is nearly constant with the shear stress, but on day 2, the low-shear stress cases have significantly higher $SC$. With a coverage that exceeds $\SI{10}{\percent}$ after three days, interactions at the substratum between neighbouring colonies such as observed by \citep{Liu.2017} become much more likely. These interactions can influence the further spreading of either colony, thus complicating the relationship between substratum coverage and time. Additionally, a small number of colonies detaches between some measurements, a process which is likely to become more prevalent as time passes.

Overall, our data demonstrates that 
the shape and size of the structures as well as the substratum coverage are sensitive to the shear stress. 
Our work additionally highlights that, as previously shown in turbulent flows \citep{Stoodley.1998}, streamers can form 
on edges of the biofilm base structures in canonical laminar channel flows.
In the following section, we further quantify the relationship between biofilm growth and shear stress and explain our observations with a simple model.

\subsection{Analysis of the relationship between biovolume and shear stress}
\begin{figure}
    \begin{subfigure}[t]{0.49\textwidth}
        \includegraphics[width=\textwidth]{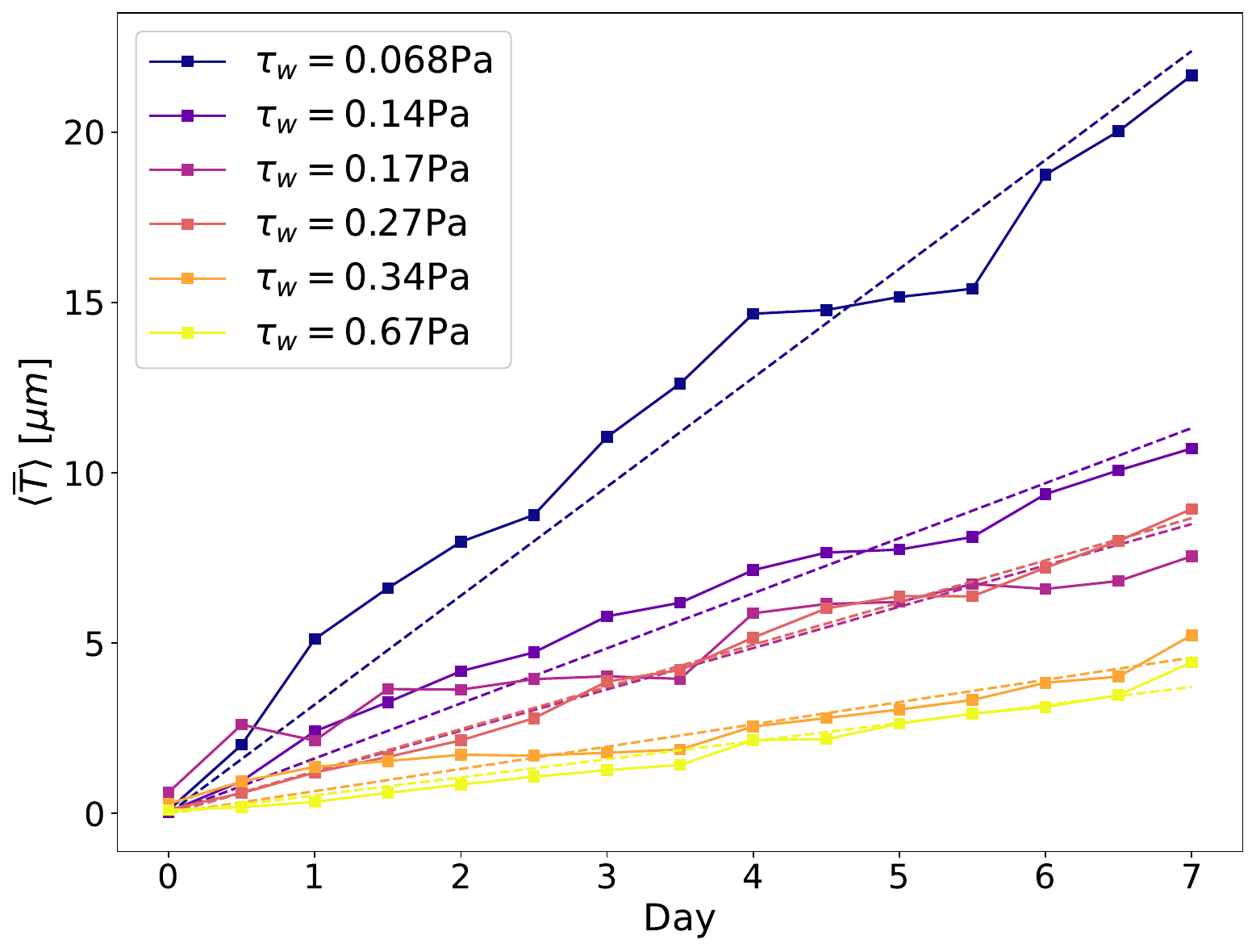}
        \caption{\textbf{Mean biofilm thickness over time.} Dashed lines represent linear least-squares fits.}
        \label{fig:meanHeightUnscaled}
    \end{subfigure}
    \begin{subfigure}[t]{0.49\textwidth}
        \includegraphics[width=\textwidth]{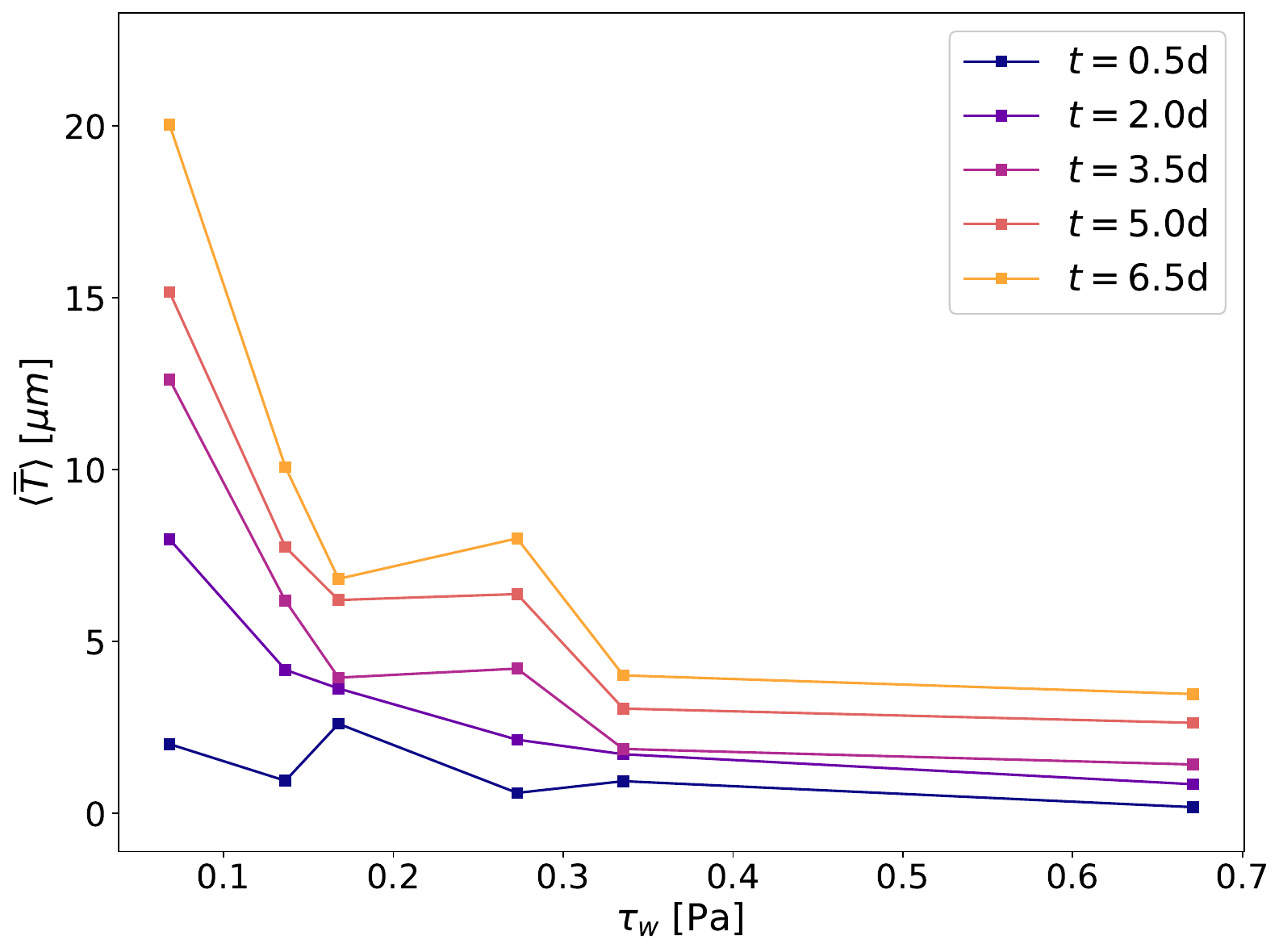}
        \caption{\textbf{Mean biofilm thickness over wall shear stress.} The values correspond to the mean value of all replicates. Data is shown for five regularly spaced measurements.}
        \label{fig:meanHeightOverTau}
    \end{subfigure}
    \caption{Change in biovolume}
    \label{fig:biovolumeData}
\end{figure}

The accumulation of biofilm can be measured by the volume of the biofilm as defined in Equation \ref{eq:biovolume}. Here, we use the mean biofilm thickness $\overline{T}$ (Equation \ref{eq:normalized_thickness}) as a normalised measure of the biovolume. This accounts for differences in the field of view that result from reflections and noisy regions within the measurements. 
Figure \ref{fig:meanHeightUnscaled} shows the development of the mean biofilm thickness with time, averaged over replicates of the same condition. We find a continuous growth of the biofilm and that the accumulation of biofilm strongly depends on the wall shear stress.
With increasing wall shear stress, the growth rate decreases. The averaged measurements reveal a clear trend. 
The mean biofilm thickness grows approximately linearly in time. 
Similar growth behaviour in time has been reported previously \citep{Horn.1997, Bakke.2001}. Gierl et al. \citep{Gierl.2020} also measured an almost linear accumulation of \textit{Bacillus subtilis} biofilm at a wall shear stress close to $\tau_w = \SI{0.068}{\pascal}$, which is the lowest wall shear stress under consideration here.

\begin{figure}
    \centering
   \includegraphics[width=0.5\textwidth]{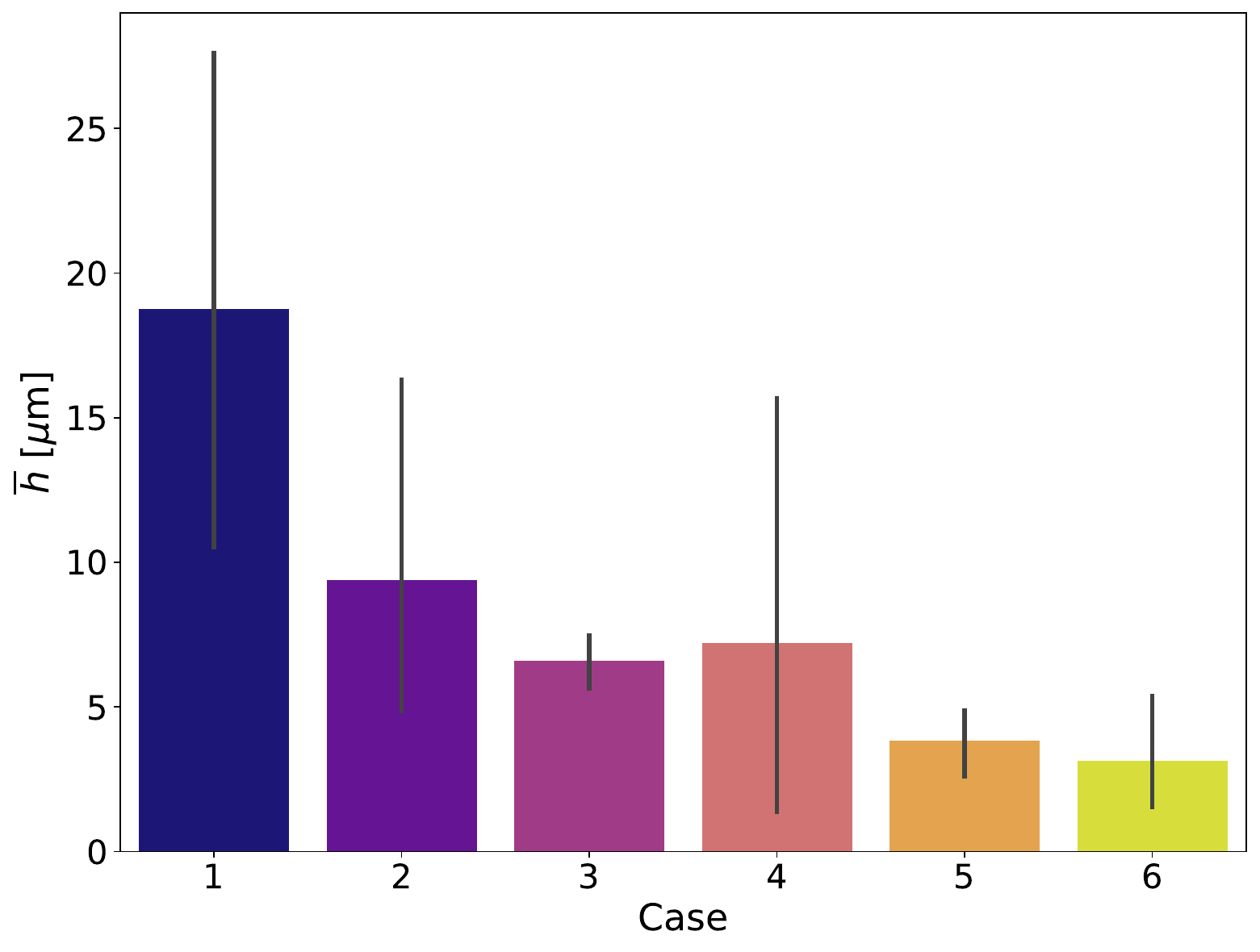}
   \caption{\textbf{Mean biofilm thickness on day six.} The error bars represent the $\SI{95}{\percent}$ confidence interval}
   \label{fig:reproducibility-biovol}
\end{figure}

Figure \ref{fig:meanHeightOverTau} shows the mean biofilm thickness as a function of the wall shear stress. Here, a clear trend of decreasing biovolume with increasing wall shear stress is visible at all times. The mean biofilm thickness after six days, along with the $\SI{95}{\percent}$ confidence intervals is shown in Figure \ref{fig:reproducibility-biovol}. 
The increase in mean biofilm thickness at $\tau_w = \SI{0.27}{\pascal}$ (case 4) again stems from an increased biofilm growth in one channel, more than five times as large as the mean of the other channels, that dominates the other replicates. The decay of the biovolume with the wall shear stress confirms the often-observed trend that increasing wall shear stress impedes biofilm development \citep{Kim.2013, Chun.2022}. 
The work by \citep{Chang.2020} measured the production of \textit{Bacillus sp.} biofilm after three days in conditions similar to our study. They reported that an increase in the wall shear stress from $\SI{0.23}{\pascal}$ to $\SI{0.68}{\pascal}$ caused an increase in the average biofilm thickness. The variance of their measurements at these two conditions, however, may explain most of this effect as the standard deviations are close to the mean values.

\subsubsection*{Friction-limited growth}
\begin{figure}
    \centering
    \includegraphics[width=0.5\textwidth]{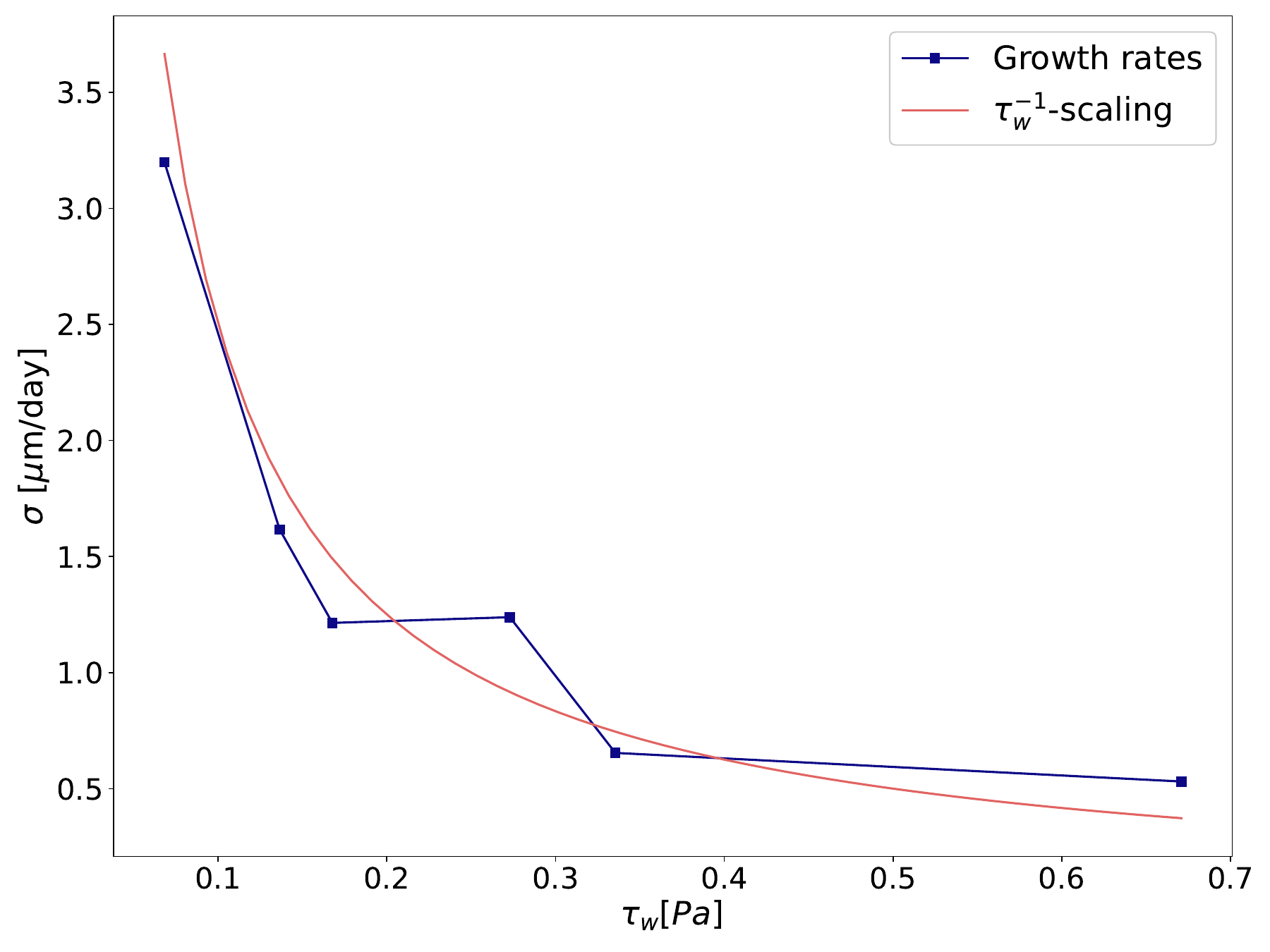}
    \caption{\textbf{Growth rate dependency on wall shear stress.} The orange line represents an inverse scaling with $\tau_w$.}
    \label{fig:growthRates}
\end{figure}

Assuming a linear growth in time, the accumulation of biofilm in Figure \ref{fig:meanHeightUnscaled} may be described using a volumetric growth rate $\sigma$,
\begin{equation}
    \overline{T}(t,\tau_w) \approx \sigma(\tau_w)t.
\end{equation}
We estimate the growth rate at each $\tau_w$ via a least squares fit using the data in Figure \ref{fig:meanHeightUnscaled}. The growth rate scales inversely proportional with the wall-shear stress, as shown in Figure \ref{fig:growthRates}. Similar growth dependence has been observed for other types of bacteria \citep{Chun.2022}. 
We additionally observe (Figure \ref{fig:timeSeries}) that the individual microcolonies mature, i.e. reach their final height, within a few days. Thereafter, the volume of each microcolony remains almost constant. 

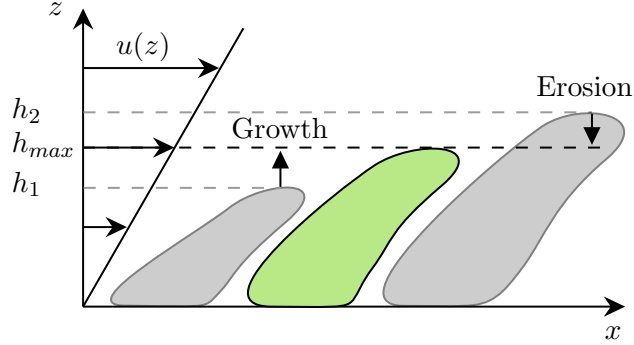
\begin{figure}
    \centering
    \tikzset{every picture/.style={line width=0.75pt}} 

\begin{tikzpicture}[x=0.75pt,y=0.75pt,yscale=-1,xscale=1]

\draw    (32,159.33) -- (299.55,159.33) ;
\draw [shift={(302.55,159.33)}, rotate = 180] [fill={rgb, 255:red, 0; green, 0; blue, 0 }  ][line width=0.08]  [draw opacity=0] (10.72,-5.15) -- (0,0) -- (10.72,5.15) -- (7.12,0) -- cycle    ;
\draw    (32,159.33) -- (32,12.33) ;
\draw [shift={(32,9.33)}, rotate = 90] [fill={rgb, 255:red, 0; green, 0; blue, 0 }  ][line width=0.08]  [draw opacity=0] (10.72,-5.15) -- (0,0) -- (10.72,5.15) -- (7.12,0) -- cycle    ;
\draw  [color={rgb, 255:red, 0; green, 0; blue, 0 }  ,draw opacity=0.5 ][fill={rgb, 255:red, 155; green, 155; blue, 155 }  ,fill opacity=0.5 ] (105.02,109.55) .. controls (122.6,95.8) and (159.72,94.07) .. (133.19,116.8) .. controls (106.65,139.53) and (108.1,143.3) .. (101.02,151.05) .. controls (93.94,158.8) and (95.27,159.3) .. (77.75,159.08) .. controls (60.23,158.87) and (78.27,159.3) .. (52.27,158.8) .. controls (26.27,158.3) and (87.44,123.3) .. (105.02,109.55) -- cycle ;
\draw    (32,119.33) -- (50.87,119.33) ;
\draw [shift={(53.87,119.33)}, rotate = 180] [fill={rgb, 255:red, 0; green, 0; blue, 0 }  ][line width=0.08]  [draw opacity=0] (10.72,-5.15) -- (0,0) -- (10.72,5.15) -- (7.12,0) -- cycle    ;
\draw    (32,159.33) -- (112,19.33) ;
\draw    (32,79.33) -- (74.53,79.33) ;
\draw [shift={(77.53,79.33)}, rotate = 180] [fill={rgb, 255:red, 0; green, 0; blue, 0 }  ][line width=0.08]  [draw opacity=0] (10.72,-5.15) -- (0,0) -- (10.72,5.15) -- (7.12,0) -- cycle    ;
\draw    (32,39.33) -- (96.87,39.33) ;
\draw [shift={(99.87,39.33)}, rotate = 180] [fill={rgb, 255:red, 0; green, 0; blue, 0 }  ][line width=0.08]  [draw opacity=0] (10.72,-5.15) -- (0,0) -- (10.72,5.15) -- (7.12,0) -- cycle    ;
\draw [color={rgb, 255:red, 155; green, 155; blue, 155 }  ,draw opacity=1 ] [dash pattern={on 4.5pt off 4.5pt}]  (32.2,61.6) -- (286,61.6) ;
\draw  [color={rgb, 255:red, 0; green, 0; blue, 0 }  ,draw opacity=0.5 ][fill={rgb, 255:red, 155; green, 155; blue, 155 }  ,fill opacity=0.5 ] (261.4,68.53) .. controls (280.87,55.6) and (319.6,62.47) .. (293.07,85.2) .. controls (266.54,107.93) and (262.64,124.23) .. (251.87,138.93) .. controls (241.1,153.63) and (236.1,158.63) .. (221.87,158.93) .. controls (207.64,159.23) and (231.4,159.2) .. (191.87,158.93) .. controls (152.34,158.67) and (241.94,81.47) .. (261.4,68.53) -- cycle ;
\draw [color={rgb, 255:red, 155; green, 155; blue, 155 }  ,draw opacity=1 ] [dash pattern={on 4.5pt off 4.5pt}]  (32.28,99.6) -- (130.5,99.6) ;
\draw    (130.5,99.08) -- (130.5,83.92) ;
\draw [shift={(130.5,80.92)}, rotate = 90] [fill={rgb, 255:red, 0; green, 0; blue, 0 }  ][line width=0.08]  [draw opacity=0] (8.93,-4.29) -- (0,0) -- (8.93,4.29) -- cycle    ;
\draw    (286,62.33) -- (286,75) ;
\draw [shift={(286,78)}, rotate = 270] [fill={rgb, 255:red, 0; green, 0; blue, 0 }  ][line width=0.08]  [draw opacity=0] (8.93,-4.29) -- (0,0) -- (8.93,4.29) -- cycle    ;
\draw  [color={rgb, 255:red, 0; green, 0; blue, 0 }  ,draw opacity=1 ][fill={rgb, 255:red, 184; green, 233; blue, 134 }  ,fill opacity=1 ] (177.35,88.36) .. controls (196.82,75.43) and (237.33,75.93) .. (210.8,98.66) .. controls (184.27,121.4) and (187.07,125.46) .. (176.3,140.16) .. controls (165.53,154.86) and (170.23,159.03) .. (156,159.33) .. controls (141.77,159.63) and (165.53,159.6) .. (126,159.33) .. controls (86.47,159.07) and (157.88,101.3) .. (177.35,88.36) -- cycle ;
\draw  [dash pattern={on 4.5pt off 4.5pt}]  (32,79.33) -- (290,79.33) ;

\draw (291,168) node [anchor=north west][inner sep=0.75pt]   [align=left] {$x$};
\draw (13,5) node [anchor=north west][inner sep=0.75pt]   [align=left] {$z$};
\draw (-5,53) node [anchor=north west][inner sep=0.75pt]   [align=left] {$h_{2}$};
\draw (47,20) node [anchor=north west][inner sep=0.75pt]   [align=left] {$u(z)$};
\draw (-5,90) node [anchor=north west][inner sep=0.75pt]   [align=left] {$h_{1}$};
\draw (104.5,63) node [anchor=north west][inner sep=0.75pt]   [align=left] {Growth};
\draw (256.75,42) node [anchor=north west][inner sep=0.75pt]   [align=left] {Erosion};
\draw (-5,70) node [anchor=north west][inner sep=0.75pt]   [align=left] {$h_{max}$};

\end{tikzpicture}
    \caption{\textbf{Three microcolonies are exposed to a shear flow.} The small colony is growing, while the larger colony is dominated by erosion. The green colony is in equilibrium.}
    \label{fig:microcolonyShear}
\end{figure}

The volume of a microcolony is approximated by $V_{mc}\approx A_{fp}h_{mc}$, where $A_{fp}$ is the footprint area and $h_{mc}$ is the height of the microcolony, respectively. We thus assume that a microcolony has a constant  cross-sectional area and formulate a balance equation for its height, 
\begin{equation}
   \frac{ \mathrm{d}h_{mc}}{ \mathrm{d}t} = g - e,
    \label{eq:balance}
\end{equation}
where $g$ is the change in height due to growth, and $e$ is the height loss rate due to erosion. If $g>e$, the microcolony grows, and if $g<e$ erosion will reduce the height of the colony. This is visualised in figure \ref{fig:microcolonyShear}.
When the right-hand side is zero ($g=e$), the microcolony has reached an equilibrium height $h_{\max}$, which is represented by the central, green structure.
We model $g$ as a constant with a value determined by the biological properties of the biofilm. In general, the growth rate can be modelled with a Monod term to account for the effects of nutrient gradients. However, here we want to focus on friction-limited growth and thus assume that the flow is fully mixed and saturated with nutrients and oxygen.

We assume that erosion depends on the imposed shear stress from the fluid and on the height of the microcolony. The simplest model of such an erosion mechanism is,
\begin{equation}
    e = \frac{C}{\mu_{mc}} \tau_{w} h
    \label{eq:erosion-constitutive relation}
\end{equation}
where $C$ is a dimensionless coefficient that depends on the biological and material properties of the biofilm and $\mu_{mc}$ is an effective viscosity of the biofilm. The dependence on the height in Equation \ref{eq:erosion-constitutive relation} is due to the fact that taller microcolonies are exposed to higher flow velocities. Therefore, the local viscous force at the surface of the microcolony, which is effectively causing the erosion, will also increase with the colony height.

By inserting \eqref{eq:erosion-constitutive relation} into the balance equation \eqref{eq:balance} and assuming an equilibrium state ($ \mathrm{d}h_{mc}/ \mathrm{d}t=0)$, we obtain the maximum height that a microcolony can reach:
\begin{equation}
    h_{max} = \frac{g \mu_{mc}}{C}\frac{1}{\tau_w}. \label{eq:hmax}
\end{equation}
We thus note that a height (or equivalently mass) balance combined with a linear constitutive relation for the erosion results in an inverse relationship between the equilibrium height of the microcolony and the imposed shear stress.

As described in Section \ref{sec:morphology}, the measured biofilms consist of many microcolonies. Here, we assume that these colonies are independent of each other and appear at a constant rate $\beta$, such that the number of microcolonies $N_{mc}$ is given by 
\begin{equation}
    N_{mc} = \beta t.
\label{eq:Nmc}
\end{equation} 
This is supported by the approximately linear growth of the substratum coverage in Figure \ref{fig:coverageUnscaled}, which corresponds to $SC = N_{mc} A_{fp} / A$ with $A_{fp}$ as the footprint area of a microcolony and $A$ as the total measurement area. Since the individual microcolonies mature quickly, we assume all colonies to be in their equilibrium state.
If each microcolony additionally has a constant and equal footprint $A_{fp}$, the collective biofilm volume of the microcolonies is given by 
\begin{equation}
    V_b = N_{mc} A_{fp} h_{max}.
\end{equation}
Using Equations \ref{eq:hmax} and \ref{eq:Nmc} yields
\begin{equation}
    V_b = \underbrace{\left (\beta A_{fp} \frac{g\mu_{mc}}{C}\right)}_{=K} \frac{t}{\tau_w}.
\label{eq:VbModel}
\end{equation}
The term $K$ encapsulates biological, material and geometrical features of the biofilms and needs to be determined empirically, but is independent of time and wall shear stress.
The biofilm volume can be normalised on a region of interest, or field of view, $A$, to obtain the final expression in our model
\begin{equation}
    \overline{T} = \frac{K}{A} \frac{t}{\tau_w}.
\end{equation}
The above expression explains the behaviour that we observe in Figure \ref{fig:biovolumeData}, namely linear growth in time, and inverse scaling with the wall shear stress.

\section{Conclusions}
We have conducted a study on the influence of wall-shear stress on the growth of \textit{Bacillus subtilis} biofilm in rectangular channels at low Reynolds numbers. Due to the wide-aspect ratio of the channels, it can be assumed that the external flow is a plane Poiseuille flow. Combining a model organism and canonical laminar flow enables a fundamental study of biofilm growth. We have used an experimental setup where biofilms are grown in 12 channels simultaneously at different shear stresses. In this way, we have reduced variations in growth that may be caused by disturbances in the external environment. Using optical coherence tomography, we obtained full volumetric scans of the biofilms every 12 hours over six days without disturbing the samples.

We found that biofilms can be regarded as a collection of microcolonies where each colony has a base structure in the form of a leaning pillar and a streamer in the form of a thin filament. While the shape, size and distribution of these microcolonies depend on the imposed shear stress,  the same structural features were observed for all shear stress values. To the best of our knowledge, this study is the first to report the consistent and robust formation of these microcolonies in laminar flow across an order-of-magnitude interval in shear stress. 
The time evolution of the biofilms suggests that the base of the microcolony develops first within the first 1-2 days. Once the base is formed, the streamlines are curved around the pillar and converge on the leeward side of the pillar. Previous work \cite{Rusconi.2010} has demonstrated that curved streamlines due to secondary flow near geometrical corners trigger the formation of streamers. In our case, streamers form on the leeward side close to the tip of the microcolonies, suggesting that this mechanism can be triggered not only by the geometry of the channel but also by the shape of the developing biofilm.
To observe the evolution of such microcolonies, the time window needs to be longer than hours to allow for the sequential forming of the base structure followed by a streamer but shorter than months to avoid a more complex, fully connected biofilm morphology. Indeed, we observe that with time, the microcolonies merge either through streamer growth or through the expansion of the pillars. Interestingly, the alignment of streamers of the many microcolonies results in a distinct macroscopic morphology characterised by a pattern of long, narrow, and elongated streaks. 

Our second main contribution is more quantitative and related to how the biofilm volume depends on shear stress over time.
We observed that the amount of biofilm within a channel, measured by its volume, grows approximately linearly over seven days for all the shear stress values. Furthermore, the growth rate was inversely proportional to the wall shear stress. 
This behaviour can be explained with a simple model based on formulating a mass balance of the biofilm. We assumed an equilibrium state such that the rate of erosion is balanced by the growth of the biofilm. Further, assuming a linear relationship between erosion and shear stress,  we could show that the biofilm volume at equilibrium is inversely proportional to the external shear stress. A second insight obtained by observing the biofilm growth was that the number of microcolonies increased over time. Using this fact, we could arrive at a final expression for the biofilm volume that scales with time and shear stress in agreement with our observations. 

The model provides valuable insight into a friction-limited growth mechanism. Having assumed that we do not have nutrient depletion, a surprisingly simple relationship seems to exist between fluid forces and biofilm growth of \textit{Bacillus subtilis} for intermediate times. Continuum models where the biofilm is modelled as an active matter may thus explicitly incorporate the dependence of shear stress and decouple the friction-limited growth rate from the nutrient-limited one. Another implication is that time and shear stress are decoupled in expression \ref{eq:VbModel}. This means that two biofilms grown under different shear stresses reach the same stage at different times. For example, reducing the shear stress by a factor of two results in the same biofilm volume as the original shear stress after twice the growth time. 

Our conclusions are based on one specific bacterial species assuming well-mixed conditions and for intermediate times. The observed growth mechanism is only applicable as long as new, independent structures appear. Future work will explore whether the specific form of micro-colonies and their growth mechanisms observed here also holds for other species and flow conditions.

\section*{Disclosure statement}
No potential conflict of interest was reported by the authors.

\section*{Funding}
This work was financially supported by the Swedish Research Council (VR) under grant 2020-04714 and through the Helmholtz Association programme “Materials Systems Engineering” under the topic “Adaptive and Bioinstructive Materials Systems”

\section*{ORCID}
Cornelius Wittig \url{https://orcid.org/0000-0001-7696-850X}\\
Romain Vallon \url{https://orcid.org/0000-0003-0770-787X}\\
Thomas Crouzier \url{https://orcid.org/0000-0002-1981-3736}\\
Wouter van der Wijngaart \url{https://orcid.org/0000-0001-8248-6670}\\
Harald Horn \url{https://orcid.org/0000-0002-9385-3883}\\
Shervin Bagheri \url{https://orcid.org/0000-0002-8209-1449}\\

\section*{Data availability statement}
The data that support the findings of this study are available from the corresponding author, C. W., upon reasonable request.

\bibliographystyle{tfnlm}
\bibliography{bibliography}

\newpage
\section*{Appendix - Time series data}
\FloatBarrier
\begin{figure}[h]
    \centering
    \includegraphics[width=0.85\textwidth]{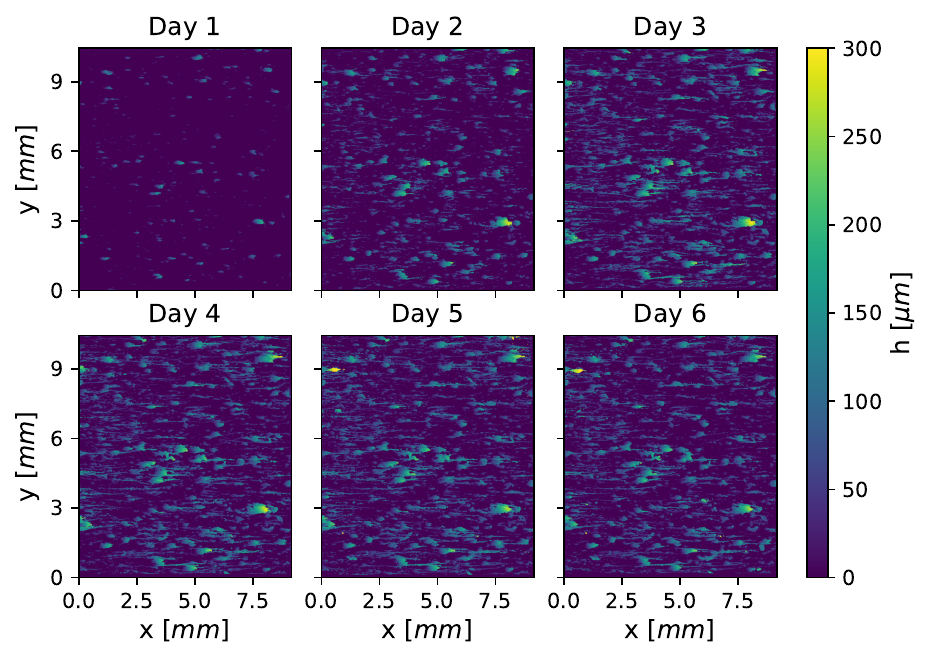}
    \caption{Case 1}
    \label{fig:timeSeries1}
\end{figure}
\begin{figure}[b!]
    \centering
    \includegraphics[width=0.85\textwidth]{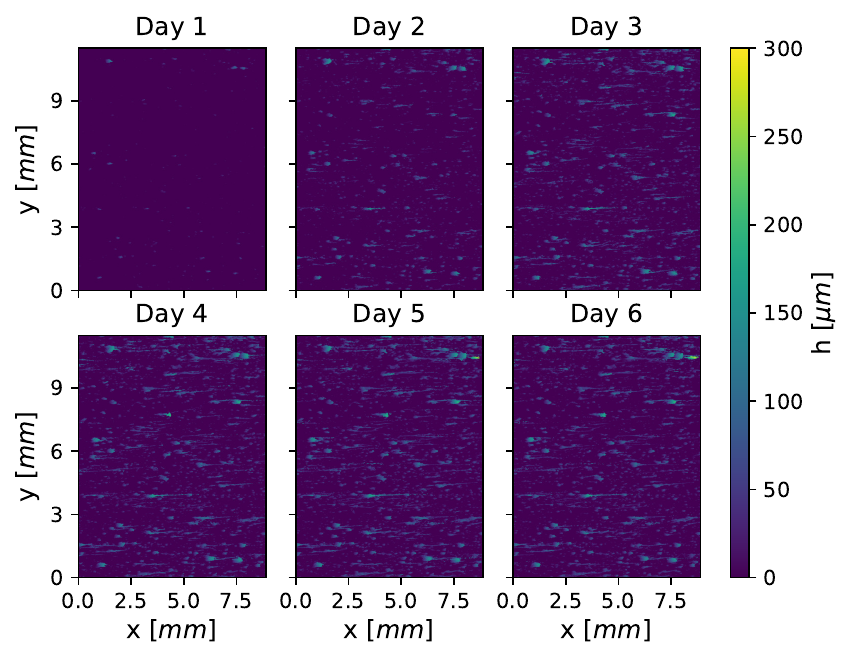}
    \caption{Case 2}
    \label{fig:timeSeries2}
\end{figure}
\begin{figure}
    \centering
    \includegraphics[width=0.95\textwidth]{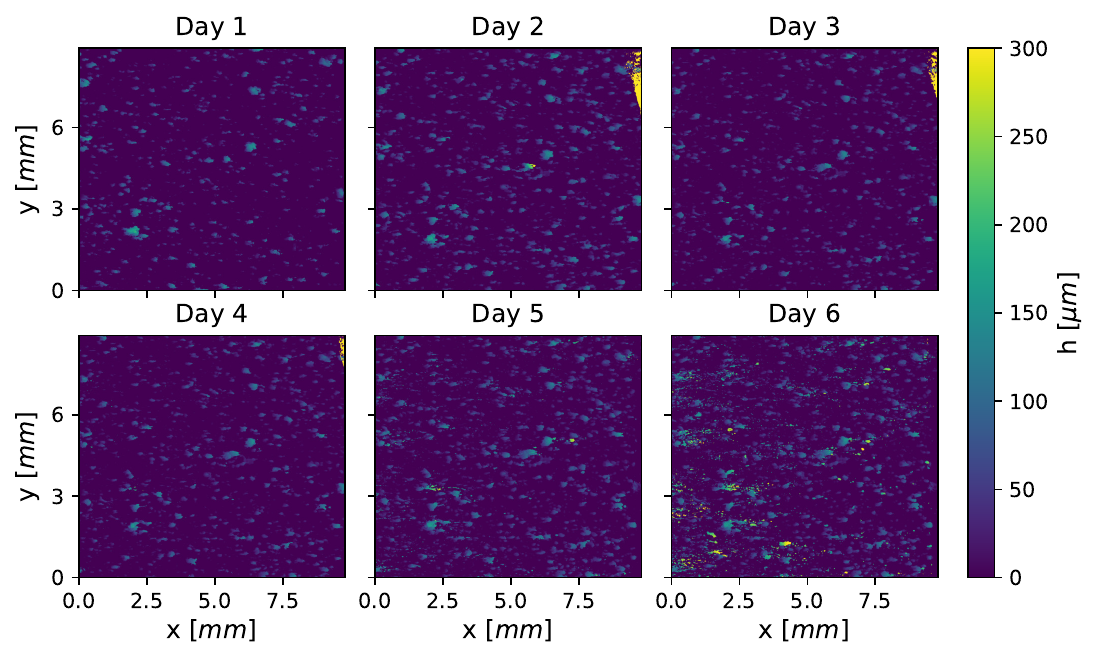}
    \caption{Case 3}
    \label{fig:timeSeries3}
\end{figure}
\begin{figure}
    \centering
    \includegraphics[width=0.95\textwidth]{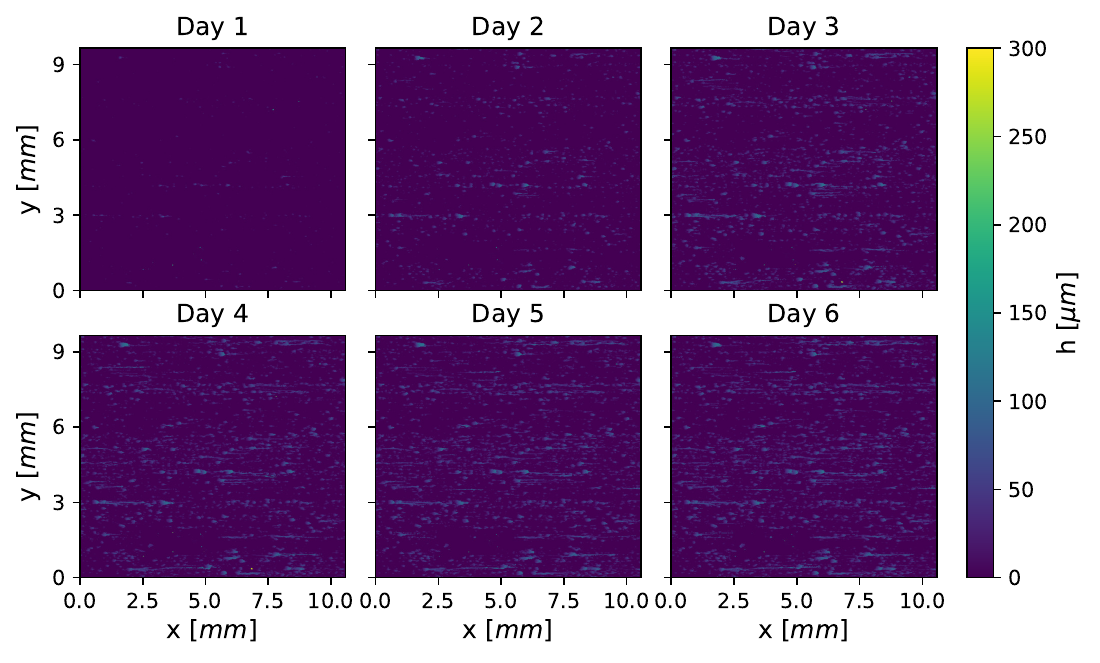}
    \caption{Case 4}
    \label{fig:timeSeries4}
\end{figure}
\begin{figure}
    \centering
    \includegraphics[width=0.95\textwidth]{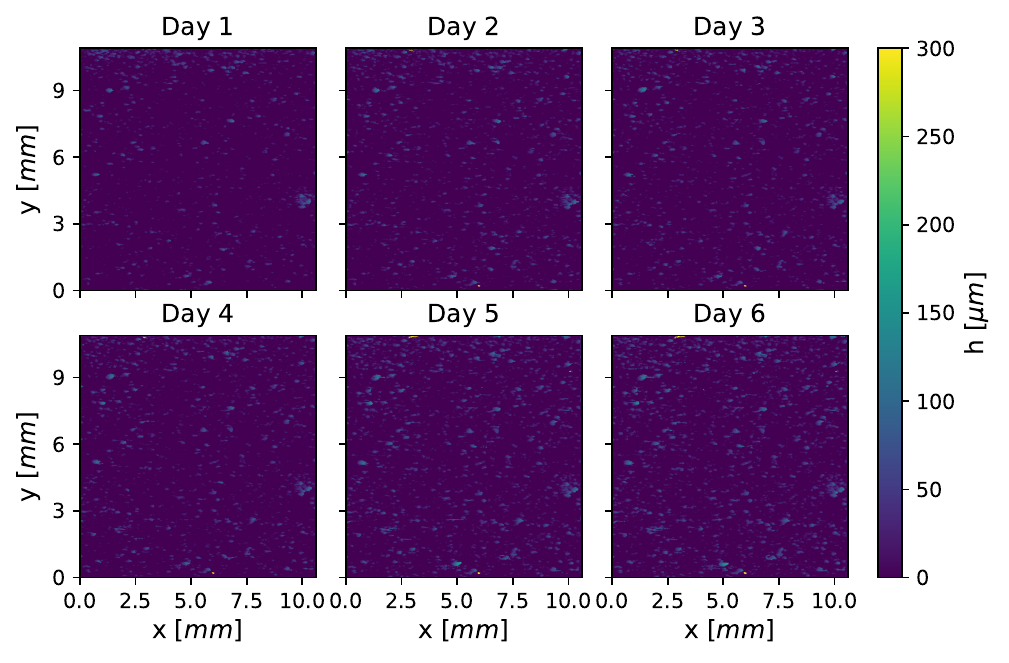}
    \caption{Case 5}
    \label{fig:timeSeries5}
\end{figure}
\begin{figure}
    \centering
    \includegraphics[width=0.95\textwidth]{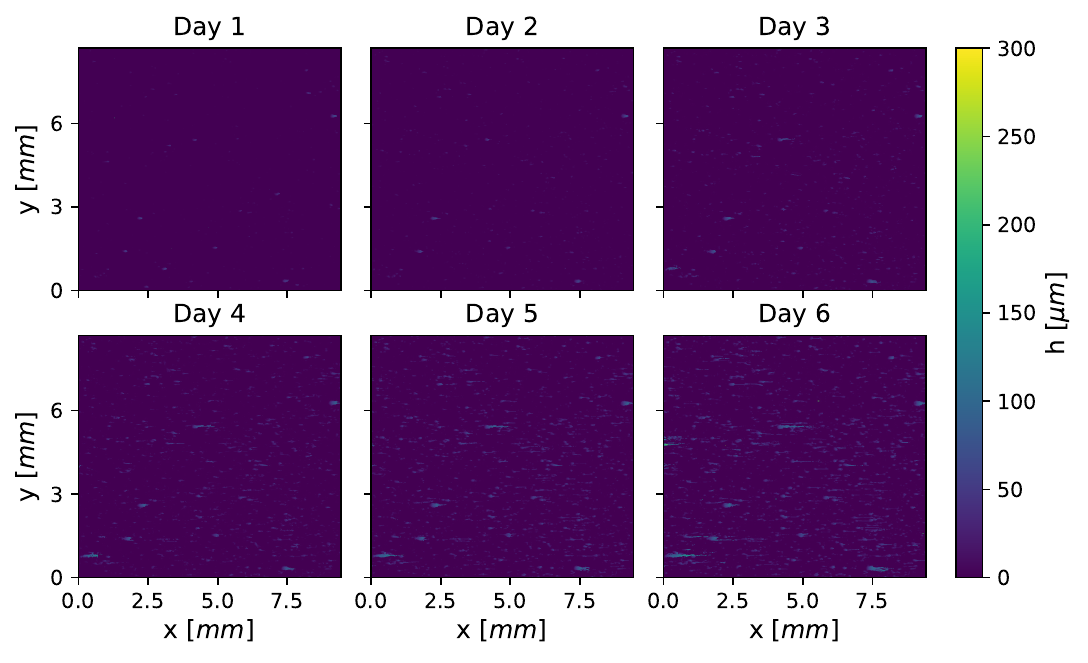}
    \caption{Case 6}
    \label{fig:timeSeries6}
\end{figure}

\end{document}